\documentclass[aps,prl,showpacs,twocolumn,superscriptaddress,longbibliography,amsmath,amssymb]{revtex4-1}
\usepackage[pdftex]{graphicx}
\usepackage{color}
\usepackage[pdftex,bookmarks,colorlinks,breaklinks]{hyperref}
\hypersetup{linkcolor=red,citecolor=blue,filecolor=dullmagenta,urlcolor=blue}
\hypersetup{pdftoolbar=true,pdfpagemode=UseNone,pdfstartview=FitH, colorlinks=true,extension=}

\usepackage{braket}
\renewcommand{\Re}{\textrm{Re}}
\renewcommand{\Im}{\textrm{Im}}

\usepackage{soul}

\begin{document}

\title{Nonlinearity-induced Scattering Zero Degeneracies for Spectral Management of Coherent Perfect Absorption in Complex Systems}

\author{Cheng-Zhen Wang}
\affiliation{Wave Transport in Complex Systems Lab, Department of Physics, Wesleyan University, Middletown, CT-06459, USA}
\author{John Guillamon}
\affiliation{Wave Transport in Complex Systems Lab, Department of Physics, Wesleyan University, Middletown, CT-06459, USA}
\author{William Tuxbury}
\affiliation{Wave Transport in Complex Systems Lab, Department of Physics, Wesleyan University, Middletown, CT-06459, USA}
\author{Ulrich Kuhl}
\affiliation{Université Côte d’Azur, CNRS, Institut de Physique de Nice (INPHYNI), 06200, Nice, France}
\author{Tsampikos Kottos}
\email[]{tkottos@wesleyan.edu}
\affiliation{Wave Transport in Complex Systems Lab, Department of Physics, Wesleyan University, Middletown, CT-06459, USA}

\begin{abstract}
We develop a Coherent Perfect Absorption (CPA) protocol
for cases where scale invariance is violated due to the presence of nonlinear mechanisms. We demonstrate,
using a microwave setting that lacks geometrical symmetries, that the nonlinearity offers new reconfigurable
modalities: the destruction or formation of nonlinear CPAs (NL-CPAs), and their frequency positioning and
bandwidth management using the incident power as a control knob. The latter occurs via the formation of
exceptional point degeneracies of the zeroes of nonlinear scattering processes. Our results establish NL-CPA
protocols as a versatile scheme for the creation of reconfigurable hot/cold-spots in complicated enclosures
(e.g. buildings or vessels) with applications to next-generation telecommunications, long-range wireless
power transfer, and electromagnetic warfare.
\end{abstract}

\maketitle

{\bf Introduction - } The wealth of the underlying mathematical structures of non-Hermitian operators has revealed
a variety of new concepts, including loss-induced transparency \cite{GSDMVASC09} unidirectional invisibility
\cite{LREKCC11,RBMOCP12,FXFLOACS13,FSA15}, hypersensitive sensing \cite{W14,HHWGGCK17,COOZWY17,KCETK22,SFKK23},
self-protecting power limiters \cite{MKV15,RKTBSAVWCKC23}, chiral state transfer \cite{DMBKGLMRMR16,NLLSHZRLCK22},
enhanced power emission \cite{GKKT23} and more (see reviews \cite{MA19,el2018non,ORNY19}). Among the various
developments, the coherent perfect absorption (CPA) protocol has attracted considerable attention. It aims to design
coherently injected multiport wavefronts that appropriately interfere to enforce perfect absorption from an absorption
center that is embedded inside a cavity -- even in cases where the losses are very weak \cite{chong2010coherent,
wan2011time,baranov2017coherent}.

Formally speaking, a CPA wavefront is associated with the eigenvector of the scattering matrix that corresponds
to a zero eigenvalue. Such zero scattering mode describes the solution of a wave operator with purely incoming
boundary conditions. When these zeroes occur on the real frequency axis, they define incident coherent wavefronts
that are completely absorbed by a lossy cavity. This is the concept of CPA \cite{chong2010coherent,krasnok2019anomalies},
that has been experimentally tested in a variety of platforms ranging from acoustics \cite{romero2016perfect},
RF \cite{schindler2012symmetric} and microwaves \cite{chen2020perfect} to optics \cite{wan2011time}. A new kind
of CPAs can emerge when two (or more) purely incoming solutions of the wave operator coalesce forming an exceptional
point degeneracy (EPD); thus leading to an anomalous line-shape of the absorbance \cite{sweeney2019perfectly,
wang2021coherent}. For example, in the case of two coalescing CPAs, the frequency dependence of the absorbance
around its maximum has a quartic (not Lorentzian!) line-shape resulting in a broadband absorption
\cite{sweeney2019perfectly,wang2021coherent,SZMRGO22}.

Most of the CPA demonstrations \cite{baranov2017coherent} and their EPDs involved simple cavity geometries
\cite{wang2021coherent,SZMRGO22}. Recently, however, CPAs have been demonstrated in complex environments
\cite{chen2020perfect} where multiple scattering and the consequent interference of many photon paths
through the medium generate extraordinary complexity and sensitivity. A reason for this paradigm shift is
the realization that CPA protocols can be used in wireless communications as a means to create hot spots in
reverberate environments \cite{FAAO20,FAAO22,HYBD21}. Other experimental settings include chaotic optical
and microwave cavities, networks of coaxial cables, and disordered systems with parity symmetry \cite{li2017random,
jiang2022coherent,sol2023covert,PKBBAKR19,RRTMSHA23,ferise2022exceptional}. Importantly, all
these cases assume linear scattering processes where the superposition principle is applicable.

On the other hand, many realistic scattering scenarios involve nonlinear processes. In wireless
communications, for example, diodes are indispensable components of wireless circuits. Similarly, optical
nonlinearities emerge naturally in photonics but up to now they have been underutilized as far as perfect
absorbing protocols are concerned. This is surprising since the incident power might offer additional
freedom for externally controlling the formation of CPAs via a self-induced variation of the impedance
of a medium. The few NL-CPA studies are either theoretical \cite{reddy2013light,
konotop18}, or they have been performed using very simple settings \cite{MSBJBLZKO18,SYRMKK22} that do
not represent typical complex scattering processes occurring in realistic scenarios, e.g., wireless
communications in reverberant cavities. Likewise, EPDs (of any kind!) are barely discussed in nonlinear
frameworks.

In this Letter, we develop a theory for the implementation of nonlinear CPAs (NL-CPAs) in complex systems
that, in general, lack geometrical symmetries. We demonstrate,
using a microwave complex network of coaxial cables,
that scale-invariance offers new modalities: external tunability of the frequencies, and linewidths (narrow
vs broad) of the CPA phenomenon. Specifically, the electrical permitivities of nonlinear elements are
externally tuned by the incident power levels; thus allowing us to control the formation of NL-CPAs and even
enforcing their coalescence. Although in many cases this coalescence occurs among complex zeroes, their
proximity to the real frequency axis constitutes these degeneracies practically indistinguishable from the
EPD-CPAs; thus allowing us to witness a broadening of the absorbance which maintains the quartic lineshape
as in the case of EPD-CPAs.

\begin{figure}
\centering
\includegraphics[width=0.9\linewidth]{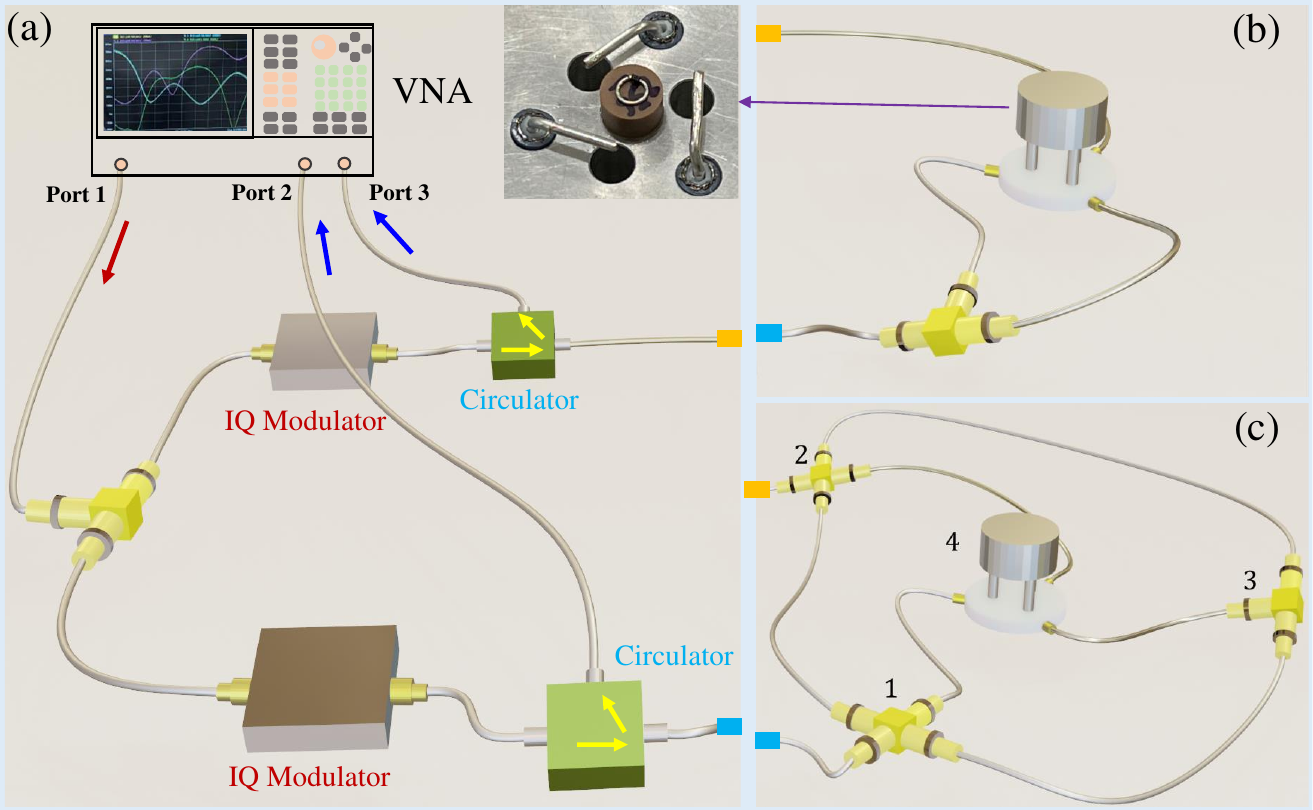}
\caption{ \label{Ring_Graph}
(a) The input signal from port 1 of vector network analyzer (VNA) is splitted into two signals by a T-
junction and then modulated by IQ modulators both in amplitude and phase. The two circulators direct the
input signals into the microwave network and the output signals into ports 2 and 3 for measurement. (b)
Ring network. (c) Tetrahedron network. The network is coupled via kink antennas to a resonator
which, in turn, is inductively coupled to a ring antenna which includes a short-circuited
nonlinear diode (see inset of (a)).
}
\end{figure}

{\bf NL-CPA protocol --} We assume that the nonlinear
mechanism is local, i.e., it depends on the magnitude of the scattering field $|\psi_{n_0}|\equiv |\langle
n_0|\psi \rangle|$ at a specific position $|n_0\rangle$ associated with the position of a nonlinear defect
(we use Dirac notation and represent all operators in a local orthonormal basis $\{|n\rangle\}$). We do
not impose any constraints on the functional form of the
nonlinearity which we generically represent as $\nu = \nu\left( \left| \psi_{n_{0}}\right|^{2} \right)$. By
neglecting excitation of higher harmonics, the wave operator may be represented in Fourier space by the
following system of coupled nonlinear algebraic equations,
\begin{equation} \label{eq1}
\left( H(\omega) + \nu\left( \left| \psi_{n_{0}}\right|^{2} \right)\left| n_{0} \right\rangle\left\langle n_{0}
\right| \right)\left| \psi \right\rangle = iD\left| s^{+} \right\rangle\,,
\end{equation}
where $H(\omega)$ is an $N\times N$ matrix that describes the dynamics inside the scattering
system in the absence of nonlinear mechanisms, the $N\times L_0$ matrix $D$ describes the coupling of the
$L_0$ transmission lines (TL) with the system, and $|s^{+}\rangle$ is the $L_0-$dimensional vector that
describes the incident wavefront in a flux-normalized basis. Wave-continuity requires $\left| s^{-} \right\rangle
= D^{T}\left| \psi \right\rangle + K\left| s^{+}\right\rangle$, where $K= K^{T}$ is the $L_0\times L_0$ direct
scattering matrix satisfying unitarity $K K^{\dagger}=1$. Imposing zero-outgoing scattering
conditions, $\left| s^{-} \right\rangle = 0$, enables us to re-express Eq.~(\ref{eq1}) as,
\begin{equation} \label{eq2}
{\tilde M}\left| \psi \right\rangle = 0; \quad {\tilde M}=\left( M + \nu\left( \left| \psi_{n_{0}}\right|^{2}
\right)\left| n_{0} \right\rangle\left\langle n_{0} \right| \right),
\end{equation}
where  $M = H(\omega) + iDK^{\dagger}D^{T}$ (see SM \cite{SM}).
We introduce the $(N - 1) \times N$ filtering matrix, $F$ with its defining properties, $F^{T}F +
\left| n_{0} \right\rangle\left\langle n_{0} \right| = \mathbb{I}_{N}, FF^{T} = \mathbb{I}_{N - 1}$ and
$F\left| n_{0} \right\rangle = 0$ to decouple the nonlinear element from the remaining $|n\rangle, n \neq
n_{0}$. The NL-CPA algorithm consists of the following steps:

{\it Step 1:} Insert the identity matrix related to $F$ above into Eq.~(\ref{eq2}) and isolate the nonlinear
terms,
\begin{equation} \label{eq3}
MF^{T}F\left| \psi \right\rangle = - \psi_{n_{0}}\left( M\left| n_{0} \right\rangle + \nu\left( \left|
\psi_{n_{0}}\right|^{2} \right)\left| n_{0} \right\rangle \right)\,.
\end{equation}

{\it Step 2a:} Apply $F$ to Eq.~(\ref{eq3}) and invert to solve for the field on the linear elements in
terms of the nonlinear contributions,
\begin{equation} \label{eq4}
F\left| \psi \right\rangle = - \psi_{n_{0}}\left( FMF^{T} \right)^{- 1}FM\left| n_{0} \right\rangle\,.
\end{equation}

{\it Step 2b:} Apply $\left\langle n_{0} \right|$ to Eq.~(\ref{eq3}),
to isolate the nonlinear part of the scattering field,
\begin{equation} \label{eq5}
\widetilde{M}_{n_0n_0} \psi_{n_0} = \left(\nu(|\psi_{n_0}|^2)+\alpha_1\right) \psi_{n_0} =
-\langle n_0 | M F^T F |\psi\rangle = \alpha_{0} \psi_{n_{0}} \,
\end{equation}
where in the third equality we have inserted Eq.~(\ref{eq4}). Above, we have introduced
the variables $\alpha_{1}\equiv \langle n_{0}|M\left| n_{0} \right\rangle$ and $\alpha_{0}
\equiv \left\langle n_{0} \right|MF^{T} \left( FMF^{T} \right)^{- 1}FM\left| n_{0} \right\rangle$
(involving only linear modeling elements).

\begin{figure}
\centering
\includegraphics[width=1.0\linewidth]{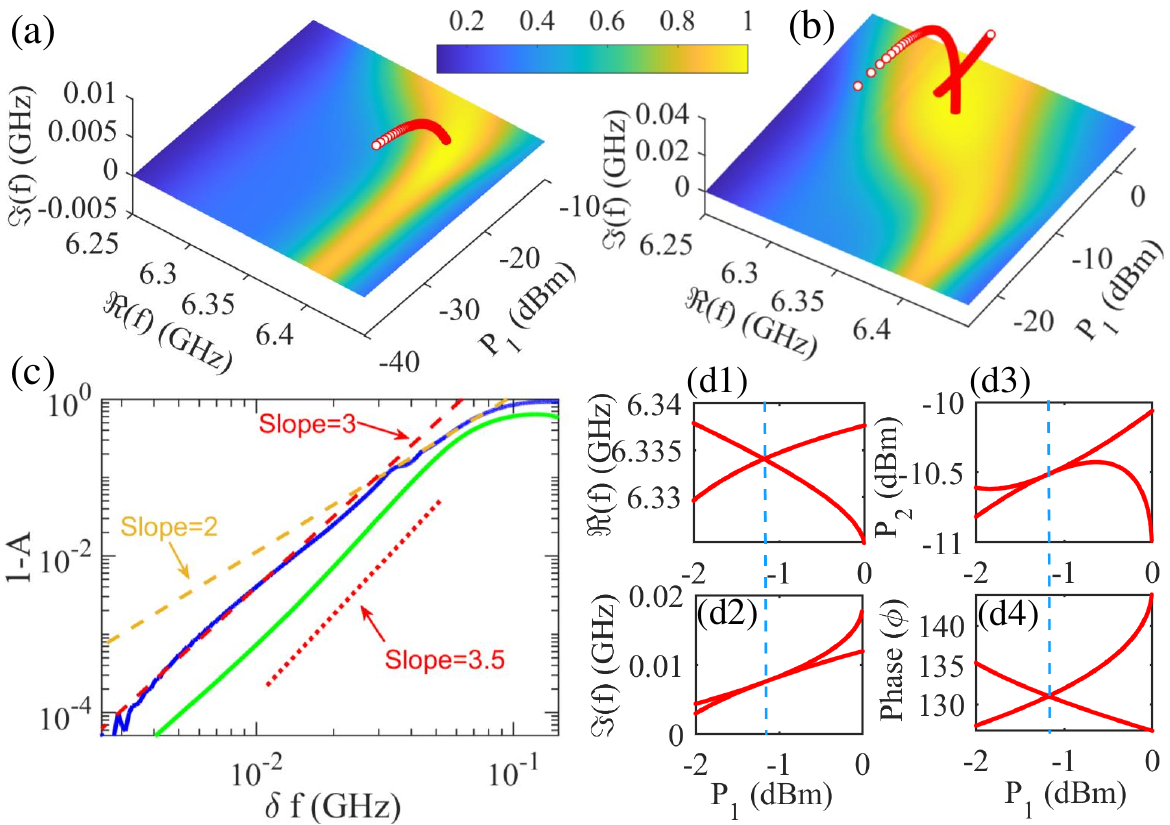}
\caption{ \label{Ring_NL_Absorption}
(a) Absorption measurements (color scale) versus frequency $\Re(f)$ and power $P_1$. Red circles represent
calculated complex zeros of the nonlinear scattering process as a function of input power $P_1$. (b) The same plot as in (a)
for broadband absorbance.
The theoretical analysis of complex zeroes (red circles) indicates the existence
of two nearby NL-CPAs that bifurcate from a complex-zero EPD occurring in proximity to the real frequency
axis. (c) The relative output signal ${\cal O}=1-A$ versus the frequency
detuning $\delta f$ from the maximum absorption conditions corresponding to case (b). The blue (green) line
is the experiment (theory). The red dashed (dotted) line is the best fit with slope 3
(slope 3.5) approximating a quartic lineshape expected for an EPD-CPA. The orange dashed line corresponds
to a Lorentzian lineshape (slope 2) and is drawn for comparison. (d) Parametric evolution of the complex
zero modes, versus the input power $P_1$: (d1) the real part $\Re(f)$ of complex zero frequency;
(d2) its imaginary part $\Im(f)$ (the small values signify proximity to the real frequency
axis); (d3) input power $P_2$; and (d4) relative phase $\phi$. The degeneracy of the complex frequencies and
the corresponding modes occurs at the same value of
$P_1$ (vertical blue dashed line) signifying a complex zero NL-EPD.
}
\end{figure}

{\it Step 3:} Solve for $\left| \psi_{n_{0}} \right|^{2}\neq 0$ given the functional form of the
nonlinearity,
\begin{equation} \label{eq6}
\left| \psi_{n_{0}} \right|^{2} = \nu^{- 1}\left( \alpha_{0} - \alpha_{1} \right)\,.
\end{equation}
A critical condition for the physicality of the solution is that $\left| \psi_{n_{0}} \right|^{2} \in
\mathbb{R}^{+}$, identified by a single parameter sweep in $\omega$, as opposed to the unaided $2L_0$
parameter sweep in relative amplitudes and phases of $\left| s^{+} \right\rangle$ and $\omega$.
Between Eq.~(\ref{eq4}) and Eq.~(\ref{eq6}) the CPA scattering field has been obtained up to an arbitrary
global phase, \(\left| \psi_{NL - CPA} \right\rangle = F^{T}F\left| \psi \right\rangle + \psi_{n_{0}}\left|
n_{0} \right\rangle\). The continuity relation and zero-outgoing scattering conditions,
immediately provides the input wave, \(\left| s_{NL - CPA}^{+} \right\rangle
= - K^{\dagger}D^{T}\left| \psi_{NL - CPA} \right\rangle\).

{\bf Experimental setup --} The microwave network is formed by coaxial cables (bonds) connected by
$n=1,\cdots, N$ T-junctions (vertices). A local lossy (saturable) nonlinearity is incorporated at
the $N$-th vertex, by substituting the T-junction with a dielectric cylindrical resonator. The
bonds are inductively coupled to the resonator by transforming the cable ends into ``kink'' antennas.
Subsequently, this resonator is inductively coupled to a metallic ring, which is positioned above its center
and is short-circuited to a diode. The kink antennas generate an EM field that is confined within the
resonator and induces a current in the diode. High currents generated by strong fields, activate a nonlinear
response of the diode which leads to a shift in the resonator's frequency.

The excitation signal is injected into port 1 of the VNA, evenly divided into two channels by a T-junction,
and subsequently routed through IQ-modulators which control the amplitude and phase of the input signals
($I_1$, $I_2$). The shaped wavefront is directed towards two circulators which separate the input
and output signals. The latter are directed to port 2 ($O_1$) and 3 ($O_2$) of the VNA for processing. The
experimental setup is shown in Fig.~\ref{Ring_Graph}.

The efficiency of our NL-CPA protocol is evaluated by the measured absorption $A=1-{\cal O}$, where
${\cal O}\equiv\frac{|O_1|^2+|O_2|^2}{|I_1|^2+|I_2|^2}$ is the total relative output signal. In
our experiment, we have used the VNA and the IQ modulators to tune four external parameters for the design
of NL-CPA wavefronts: the incident frequency $f\in [6.25,6.45]$\,GHz, the relative phase of the
injected field amplitudes $\phi = \phi_1 - \phi_2\in [0^{\circ},360^{\circ}]$ at channels 1 and 2 and their
corresponding input power $P_1$ (channel 1) and $P_2$ (channel 2) with $P_{1,2} \in [-45, 5]$\,dBm.

{\bf Implementation --} The theoretical modeling of the network platform and the NL-CPA protocol
implementation can be found in the SM \cite{SM}.

{\it Intensity manipulation of the CPA frequency}. We first demonstrate the nonlinear CPA modality associated
with the intensity manipulation of the CPA frequency. For this purpose, we have used the ring network (N=2)
of Fig.~\ref{Ring_Graph} (b) where this modality is accessible in the available power and frequency range
of our VNA. Figure~\ref{Ring_NL_Absorption}(a) reports the parametric evolution of the
complex zeroes of the nonlinear scattering process (red circles) versus the input power $P_1$. The NL-CPA
occurs when a complex zero crosses the real frequency axis at $f^{cpa}\approx6.397$\,GHz, with input power
$P_1^{cpa}\approx-16.10$\,dBm, relative phase $\phi^{cpa}\approx97^{\circ}$ and power ratio $R^{cpa}\equiv
P_1^{cpa}/P_2^{cpa}\approx2.9$.

The experimental investigation, has been guided by the prediction of our NL-CPA theory which allowed to
confine the large parameter space to a smaller domain. A maximum absorbance of
$A^*=99.998\%$ occurs at frequency $f^*\approx 6.391$GHz for
a wavefront with $P_1^*\approx-17.67$\,dBm, $R^*\equiv P_1^*/P_2^*\approx2.2$, and $\phi^*\approx 97^{\circ}$
(For detailed comparison between theory and experiment, see SM Figs.~S1,S2 \cite{SM}). The small discrepancies
between the predicted NL-CPA wavefront and the experimental results are associated with the measuring precision
of the electrical permittivity of the
coaxial cables and the uncertainty in the parameters used to model the lossy nonlinearity \cite{SM}.
Figure \ref{Ring_NL_Absorption}(a) reports the measured absorbance $A$ versus
frequency $f$ and injected power $P_1$ (for fixed $R^*$ and phase $\phi^*$). This example
demonstrates a narrowband $\sim20$\,MHz absorption with $A\geq 95\%$.

{\it Intensity manipulation of absorbance band-width by NL-EPDs}. An example where the injected power
levels are employed as an external parameter for reconfiguring the absorbance bandwidth is shown in
Fig.~\ref{Ring_NL_Absorption}(b). For the previous ring-graph configuration, we find a broadband $\sim 50$\,MHz
high absorbance ($A>95\%$) for increased injected powers.
The experimental parameters of the wavefront that results in a peak absorbance $A^*\approx99.55\%$ are
$f^*\approx6.34$\,GHz with $P_1^*\approx-3.5$\,dBm, $R^*\approx5.5$ and $\phi^*\approx 66^{\circ}$.
In this parameter range, the NL-CPA protocol predicted two NL-CPAs (a) at $f^{\text{cpa}}\approx6.346\,$
GHz with $P_1^{\text{cpa}}\approx-3.6$\,dBm, power ratio $R^{cpa}\approx6.3$ and relative phase
$\phi^{cpa}\approx120^{\circ}$; and (b) at $f^{\text{cpa}}\approx6.327$\,GHz with $P_1^{\text{cpa}}
\approx-2.3$\,dBm, $R^{cpa}\approx6.6$ and  $\phi^{cpa}\approx136^{\circ}$. The phase discrepancy between
the experimental and the theoretical values is associated with the precision of the extracted experimental
parameters of the network. It turns out that the wavefront phase is the most sensitive among all other
characteristics of the optimal wavefront (see SM and Figs.~S1,S2 \cite{SM}).

\begin{figure*}
\centering
\includegraphics[width=0.8\linewidth]{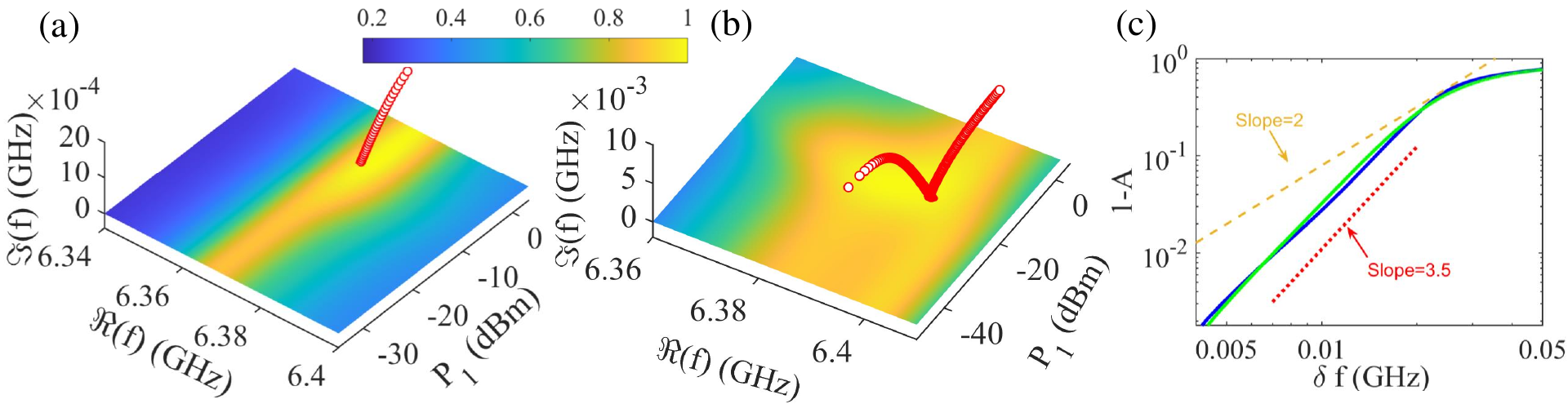}
\caption{ \label{Tetra_NL_Absorption}
Measured absorption (color scale) versus input frequency $\Re(f)$ and power $P_1$ for (a) narrowband and (b)
broadband NL-CPA. The red circles in both (a,b) are the theoretically calculated trajectories of the complex zeroes as
$P_1$ varies. In (b), two NL-CPAs, close to one-another, emanate from a complex zero EPD occuring in the proximity
of the real-frequency axis ($\Im(f^{EPD})\sim 10^{-4}$). For all practical purposes,
these two NL-CPAs form an EPD-CPA. (c) The relative output signal
(${\cal O}=1-A$) versus the frequency detuning $\delta f$ at $P_1^{cpa}$. The blue (green) line
indicates the experimental (theoretical) scaling of ${\cal O}$ versus detuning from $f^*$ ($f^{EPD}$).
The red dotted line indicates the best power-law fit with slope 3.5 (close to quartic behavior expected
for EPD-CPAs). A quadratic (Lorentzian) scaling is shown (orange dashed line) for comparison.
}
\end{figure*}

We have also analyzed the shape of the absorption spectrum in the vicinity of $f^*$. In Fig.~
\ref{Ring_NL_Absorption}(c) we report the measured relative outgoing signal ${\cal O}=1-A$ versus $\delta
f\equiv f-f^*$ (blue line) in a double-logarithmic fashion. We find a cubic power-law scaling ${\cal O}\sim
(\delta f)^3$ (see dashed red line) which is distinct from the traditional Lorentzian quadratic behavior.
The latter is shown for comparison with a dashed orange line. In the same figure, we also present the
scaling of the theoretical result (green line) and find that ${\cal O}\sim (\delta f)^{3.5}$ which is
reminiscent of the quartic scaling occurring in the case of EPD-CPAs. In fact, by tracing the theoretically
evaluated NL-CPAs in the complex frequency domain we have identified an EPD of the scattering zeroes (see
red circles in Fig. \ref{Ring_NL_Absorption}(b)) occurring at $f^{\text{EPD}}\approx (6.335+i\cdot 0.0075)$\,
GHz with $P_1^{\text{EPD}}\approx-1.18$\,dBm, $R^{EPD}\approx8.6$ and $\phi^{EPD}\approx131^{\circ}$. The
small imaginary part of the complex zeroes-EPD signifies its proximity to the real frequency axis and
justifies the fact that the experimental absorption lineshape demonstrates broadband characteristics
similar to the one predicted for an EPD-CPA \cite{sweeney2019perfectly,wang2021coherent,SZMRGO22}.

To further examine the nature of the complex-zero degeneracy, we examined in detail the parametric
evolution (versus input power $P_1$) of the complex frequency, amplitudes, and
relative phases of the complex-zero modes, see Figs.~\ref{Ring_NL_Absorption}(d1)-(d4). All these
parameters coalesce simultaneously at the same input power signifying the formation of a complex-
zero NL-EPD.

{\it Generalization to Complex Networks}. Next, we implement the NL-CPA protocol to a tetrahedron network
$(N=4)$, see Fig.~\ref{Ring_Graph} (c).
In the absence of nonlinearities, and for incommensurate bond lengths, this network shows many typical
features of wave chaos \cite{KS97,KS04,HGK18}; thus offering a platform for simulating real-world
scattering environments.

Using the NL-CPA protocol we have predicted the formation of a NL-CPA by tracing the trajectory of the
complex zeroes (red circles) of the nonlinear scattering process and identifying the point at which it
crosses the real axis. The wavefront parameters that we have found are  $(f^{cpa}, P_1^{cpa})\approx
(6.369\,GHz, -6.6\,dBm)$ for input power ratio $R^{cpa}\approx 5.96$ and relative phase $\phi^{cpa}\approx
191^{\circ}$. Again, these estimations allowed us to reduce the parameter space in our experimental
implementation. We have found that an optimal absorption of $A^*\approx99.97\%$ occurs at $f^*\approx
6.371$\,GHz with injected power $P_1^*\approx-5.09$\,dBm, power ratio $R^*\approx 6.03$ and relative phase
$\phi^* \approx 207^{\circ}$ between the two injected wavefronts. Figure~\ref{Tetra_NL_Absorption}(a)
shows the measured absorbance (color-grade) versus incident power $P_1$ and frequency $f$ for fixed relative
power ratio $R^*$ and relative phase $\phi^*$ (see SM Figs.~S3(a,b),~S4(a1-d1) \cite{SM}).

We have also identified broadband absorption scenarios in a tetrahedron network (bandwidth $\approx 20$MHz
as opposed to $\approx 6.3$MHz in the previous case, for $A>95\%$). An example is reported in Fig.~
\ref{Tetra_NL_Absorption}(b) where we show the measured absorbance (color scale) for
varying incident frequency and injected power $P_1$ (fixed relative power $R^*\approx 0.45$ and relative phase
$\phi^*\approx 301^{\circ}$). At $f^*\approx 6.39\,$GHz and $P_1^*\approx -19.05$\,dBm the absorbance
acquires its maximum value $A^*\approx99.93\%$. In comparison, the NL-CPA protocol predicted the existence
of two nearby NL-CPA wavefronts. These correspond to the wavefront parameters for which the complex zeroes
(red circles) cross the real frequency axis in Fig.~\ref{Tetra_NL_Absorption}(b) while varying $P_1$: (a)
at $f^{cpa} \approx 6.397\,$GHz with $P_1^{cpa} \approx -19.24$ dBm, $R^{cpa}\approx 0.68$, and $\phi^{cpa}
\approx 377.8^{\circ}$; and (b) at $f \approx 6.396\,$GHz with $P_1^{cpa} = -19.13$ dBm, $R^{cpa} \approx 0.71$,
and $\phi^{cpa} \approx 376.4^{\circ}$. Further analysis and comparison between theory and
experimental data is provided in the SM \cite{SM}, see Figs.~S3(c,d)
and Figs.~S4(a2-d2). The degeneracy occurs slightly above the real
plane (imaginary part of complex EPD zero $\sim 10^{-4}$) which is completely untraceable in the experiment;
thus emulating the behavior of an actual NL-CPA EPD. The exact parameters of the complex-zero NL-EPD
are $f^{EPD} \approx (6.396+0.00027i)\,$GHz, $P_1^{EPD} \approx -19\,$dBm, $R^{EPD} \approx 0.69$,and
$\phi^{EPD} \approx 376.2^{\circ}$. We have confirmed that at this point the complex-zero modes form an
EPD, by analyzing the parametric evolution of their complex frequencies, $P_2$, and relative phase $\phi$
that characterize them versus the injected power $P_1$. All of them coalesce at $P_1^{EPD}$ as expected
for an EPD (for a detailed description of the NL-EPD modes see SM Fig.~S5 \cite{SM}).
Further confirmation of the EPD was provided by analyzing the lineshape of the absorbance in the proximity
of the (quasi-)degeneracy, see Fig.~\ref{Tetra_NL_Absorption}(c). Similar to the previous case of the
ring-graph, we find that the scaling of the relative outgoing signal with respect to the detuning $\delta f$
is best fitted by a super-Lorentzian power-law  ${\cal O}\sim (\delta f)^{3.5}$ which resembles the quartic
line-shape expected for an EPD-CPA \cite{sweeney2019perfectly}.

{\bf Conclusions - } We have developed a NL-CPA protocol that is generically applicable to any scattering
setup. The scheme utilizes the nonlinear light-matter interactions as a
reconfigurable knob to tailor the features of the absorption spectrum through external
power control. We have implemented our proposal using microwave networks of coaxial cables and demonstrated
a variety of new modalities due to the nonlinearity: Control in the creation/annihilation of NL-CPAs,
their frequencies, and band-width manipulation for which high absorption occurs. The latter is achieved
in the proximity of zero-EPDs of the scattering process.
Our results establish NL-CPA protocols as a versatile scheme for the creation of hot (or cold)-
spots in reverberate environments.

{\it Acknowledgements --} We acknowledge partial funding from NSF-RINGS ECCS (2148318), and Simons
Foundation (MPS-733698).


\newpage
\begin{widetext}
\begin{center} \Large \bf
	Supplemental Material
	for manuscript: \\Nonlinearity-induced Scattering Zero Degeneracies for Spectral Management of Coherent Perfect Absorption in Complex Systems
\end{center}

\begin{center}
Cheng-Zhen Wang$^1$, John Guillamon$^{1}$, William Tuxbury$^{1}$, Ulrich Kuhl$^2$, and Tsampikos Kottos$^1$

$^1$ Wave Transport in Complex Systems Lab, Department of Physics, Wesleyan University, Middletown, CT-06459, USA\\
$^2$ Universit\'{e} C\^{o}te d'Azur, CNRS, Institut de Physique de Nice (INPHYNI), 06108 Nice, France, EU\\
\end{center}
\end{widetext}
\renewcommand\thefigure{S\arabic{figure}}
\setcounter{figure}{0}
\setcounter{page}{1}
\renewcommand{\thepage}{S\arabic{page}}
\setcounter{equation}{0}
\renewcommand\theequation{S\arabic{equation}}
\setcounter{section}{0}
\section{Theoretical Model}
\subsection{Networks}
The transport characteristics of the system are modeled by a fusion of two primary components:
the first component is a metric graph model of one dimensional wires supporting only one propagating mode connected by vertices--this model demonstrates wave chaotic scattering statistics;
the second component are the steady state solutions of a set of nonlinear coupled mode theory equations to model the nonlinear cavity mode response of the resonator.
We will consider a microwave graph with $N$ vertices (or nodes), of which $L_0$ of them are connected to transmission lines (TL). The coupling to the TL is described by the $N\times L_0$ matrix $D$ with elements $1$ ($0$) when a vertex is connected (not connected) to a TL.
For example, in the case where the 1st and 2nd vertices are connected to TL, with the $N-2$ other vertices being internal vertices, the transpose coupling matrix takes the form
\begin{equation}
\label{cmatrix}
D^T = \begin{bmatrix}
1 & 0 & \cdots & 0 \\
0 & 1 & \cdots & 0 \\
\end{bmatrix}\,.
\end{equation}

Bonds (or edges) are one-dimensional connections between vertices, where the position on each bond from vertex \(\mu\) to \(\beta\) can be defined as \(x_{\mu\beta}\), with \(x_{\mu\beta}=0\) on vertex \(\mu\) and \(x_{\mu\beta}=L_{\mu\beta}\) on vertex \(\beta\), and \(x_{\beta\mu}=L_{\mu\beta}-x_{\mu\beta}\).
To account for how the bonds connect the vertices, we will introduce the symmetric \(N\times N\) matrix $A$ which takes entries of 1 for connected vertices and 0 otherwise, with all diagonal elements set to zero as we do not allow self-connected vertices.
In addition to the \(A\) matrix, we also use another symmetric \(N\times N\) matrix, \(L\), which contains the bond lengths for the connections between the vertices.
For a simple fully connected tetrahedron graph, they take the following form:
\begin{equation}
A = \begin{bmatrix}
0 & 1 & 1 & 1 \\
1 & 0 & 1 & 1 \\
1 & 1 & 0 & 1 \\
1 & 1 & 1 & 0 \\
\end{bmatrix}
;\quad L = \begin{bmatrix}
0 & L_{12} & L_{13} & L_{14} \\
L_{21} & 0 & L_{23} & L_{24} \\
L_{31} & L_{32} & 0 & L_{34} \\
L_{41} & L_{42} & L_{43} & 0 \\
\end{bmatrix}\,.
\end{equation}
Note that \(L_{\mu\beta}=L_{\beta\mu}\). Finally, throughout the manuscript, we will be assuming that the nonlinearity is located at vertex $n_0=N$.

\subsection{Linear Wave Network Theory and CPA conditions}
First we address the wave propagation at the linear subspace of the networks. The one-dimensional bonds can model the coax cables where the microwaves propagate. The wave equation that describes the electric potential difference \(\psi_{\mu\beta}(x_{\mu,\beta})\) at position $x_{\mu,\beta}$ along the $L_{\mu,\beta}-$bond is given as
\begin{equation}
\label{A3}
\frac{d^2}{d{x_{\mu\beta}}^2}\psi_{\mu\beta}\left(x_{\mu\beta}\right)+\frac{\omega^2\epsilon}{c^2}\psi_{\mu\beta}\left(x_{\mu\beta}\right)=0\,,
\end{equation}
where $\epsilon$ is the dielectric constant of the coaxial cables, $\omega = 2\pi f$ is the angular frequency with $f$ the frequency and $c$ is the speed of light in vacuum. The wave number is $k = \sqrt{\epsilon}\omega/c$.
It is convenient to define the wave at a vertex $\mu$ as $\psi_{\mu,\beta}(x_{\mu,\beta}=0)=\psi_{\mu}=\phi_{\mu}$ where we have introduced the vertex field $\Phi = (\phi_1,\phi_2, \cdots, \phi_{N})^T$.

The solutions of the above Eq. (\ref{A3}) must be supplemented with the appropriate boundary conditions at the vertices. First, we consider wave continuity at each vertex, i.e., \(\psi_{\mu\beta}\left(x_{\mu\beta}=0\right)=\phi_{\mu}\) for any bond that connects the vertex \(\mu\) with any vertex \(\beta\). Using the vertex scattering field $\Phi$ we write the wave at each bond as
\begin{equation} \label{Eq:bondwave}
\psi_{\mu\beta}\left(x_{\mu\beta}\right)=\phi_{\mu}\frac{\sin{k\left(L_{\mu\beta}-x_{\mu\beta}\right)}}{\sin{kL_{\mu\beta}}}+\phi_{\beta}\frac{\sin{kx_{\mu\beta}}}{\sin{kL_{\mu\beta}}}\,.
\end{equation}

Next, we impose current conservation at each vertex. This condition takes the form
\begin{align} \label{Eq:current-cont}
&\sum_{\beta}{\frac{d}{d{x_{\mu\beta}}}\psi_{\mu\beta}\left(x_{\mu\beta}=0\right)+\sum_{\mu=1,2}\delta_{\mu,\alpha}\frac{d}{d{x_\mu}}\psi_{\mu}\left(x_\mu=0\right)}=0.
\end{align}

Finally, the CPA condition is satisfied by enforcing only incoming wave forms on the TLs
\begin{equation}\label{Eq:CPA-input}
\psi_\mu\left(x_\mu\right)=I_\mu\delta_{\mu\alpha}e^{-ikx_\mu}\,,
\end{equation}
where \(I_\mu\) is the amplitude of the injected wave at vertex \(\mu\) and the subscript $\alpha = 1, 2$ indicates the TL. The position $x_\mu\in [0,\infty)$ at the TL is $x_\mu=0$ at the $\mu$-vertex where the TL is connected.
Note that by excluding a counter-propagating wave \(R_\mu\delta_{\mu\alpha}
e^{ikx_\mu}\), we ensure that there are no outgoing waves (CPA-condition).

\subsection{Modeling of nonlinear resonator}\label{CMTSec}

Next, we want to evaluate the field at the edge-points of the three kink-antennas that are closer to the nonlinear resonator. To this end, we first calculate the scattering field ${\tilde a}(t)$ at the nonlinear resonator. We use a Temporal Coupled Mode Theory modeling to describe the coupling between the kink antennas and the resonator
\begin{align}
i\frac{d}{dt}\tilde{a}\left(t\right)&=\left(\Omega-\frac{i}{2}D_RD_R^T\right)\tilde{a}\left(t\right)+iD_R\tilde{S}^+\nonumber\\
\ket{\tilde{S}^-}&=K\ket{\tilde{S}^+}+D_R^T\tilde{a}(t)\,.\label{TCMT}
\end{align}
where for simplicity of the modeling we assume that $K=-\mathbb{I}_3$ (no direct coupling between the antennas) and
$D_R=(\gamma_{l_1}, \gamma_{l_2}, \gamma_{l_3})$ describes the
coupling of the resonator to the three kink antennas. Each of these antennas is connected to a corresponding $l_n$-vertex of the remaining graph. The non-linear resonant frequency $\Omega=\omega_0-\frac{1}{2}\nu(|{\tilde a}|^2)$ models the combined resonator-diode system, where $\omega_0$ is the linear angular frequency of the resonator in the absence of the coupling with the ring-diode and might also include radiative (or/and Ohmic) linear losses.

We assume that the scattering process does not excite higher harmonics. In this respect, the field at the resonator and the three antennas take the form
\begin{equation}
\tilde{a}\left(t\right)=\phi_N\cdot e^{-i\omega t};\quad\quad
\ket{\tilde{S}^\pm}=\ket{s^\pm}\cdot e^{-i\omega t}\,,
\end{equation}
where $\ket{s^{\pm}}=(s_{l_1}^{\pm},s_{l_2}^{\pm},s_{l_3}^{\pm})^T$ is a three-dimensional vector that describes the amplitudes of incoming (+) and outgoing (-) waves from the antennas to the resonator. The indices $l_n$ label the kink-antenna that is attached to the vertex $l_n$ of the graph.
Substituting the above expressions in Eqs. (\ref{TCMT})
we get the following equations that describe the steady-state scattering fields
\begin{eqnarray}
\label{sssf}
&\left(\omega-\Omega+\frac{i}{2}D_RD_R^T\right)\ \phi_N=iD_R\ket{s^+},\quad\quad\quad \label{Eq:cmt-stationary-1}\\
&\ket{s^-} = -\mathbb{I}\ket{s^+}+D_R^T\phi_N.\label{Eq:cmt-stationary-2}
\end{eqnarray}
We proceed by considering the coupling strengths between the resonator and the three kink-antennas as fitting parameters.  Using Eq. (\ref{Eq:cmt-stationary-2}), we express the wave at the origin point of the cable antennas (proximity to the resonator), as
\begin{equation} \label{A10}
\phi_N^{(l)}=s_l^+ + s_l^- = \gamma_l \phi_N,
\end{equation}
where the index $l=l_n$ labels the kink-antenna that is attached to the vertex $l$ of the graph. The field at a
generic position $x_{Nl}\in[0, L_{Nl}]$ at the cables of the kink antennas ($L_{N,l}$ are the lengths of the kink-antennas)
can be generally expressed as a combination of counter-propagating waves. We have
\begin{align}
\psi_{l,N} (x_{lN})= s_l^+ e^{-ikx_{Nl}} + s_l^-e^{ikx_{Nl}}
\label{general}
\end{align}

Using Eqs.~(\ref{A10},\ref{general}) we write the field amplitude at the position $x_{N,l}=L_{N,l}$ (endpoint/vertex of the kink-antenna that connects to the rest of the graph) as
\begin{align}\label{Eq:phi-s}
\phi_{l} = -2is_l^+\sin(kL_{Nl})+\gamma_le^{-ikL_{Nl}}\phi_N, \quad l\neq N
\end{align}
which allows us to solve for the coefficients $s_l^+$ in terms of the field amplitudes $\phi_l$. We have:
\begin{align}\label{Eq:phi-s2}
s_l^+=i\frac{\csc\left(kL_{Nl}\right)}{2} \left[\phi_l-\gamma_le^{-ikL_{Nl}}\phi_N\right]
\end{align}

\subsection{Hybrid nonlinear resonator-network modeling}

By combining Eqs.~(\ref{Eq:bondwave})-(\ref{Eq:CPA-input}) from the linear subnetwork with Eq.~(\ref{Eq:phi-s}) from the coupled mode theory analysis above, we arrive to the following set of equations
\begin{equation}
\label{hybrid}
\left(H^{(N-1)}-iFDD^TF^T\right)F\Phi = -\phi_N\cdot Y\,,
\end{equation}
where $D$ is the $N\times L_0$ graph-TL coupling matrix Eq. (\ref{cmatrix}), $F$ is a $(N-1)\times N$ filtered matrix, with elements \(F_{ij} = \delta_{i,j} \) and $M^{(N-1)}$
is the $(N-1)\times (N-1)$ matrix with elements
\begin{equation}
H_{\mu,\beta}^{(N-1)} =
\begin{cases}
- \sum_{\gamma \neq \mu} A_{\mu\gamma} \cot(kL_{\mu\gamma}) & \text{if } \mu = \beta \\
A_{\mu\beta} \csc(kL_{\mu\beta}) & \text{if } \mu \neq \beta
\end{cases}
\end{equation}
where $\mu,\beta=1,\cdots, N-1$ and $\gamma=1,\cdots,N$.
The $(N-1)$ dimensional vector $Y$ has elements $Y_\mu=\delta_{\mu,l_n}\gamma_{l_n}\csc\left({kL}_{Nl_n}\right)$.
It should be noted that Eq. (\ref{hybrid}) involves only the linear sub-network i.e. it excludes the nonlinear vertex $N$.

Furthermore, the wave amplitude $\phi_N$ at the nonlinear resonator, can be expressed in terms of the field amplitudes $\phi_{l_n}$ at the $l_n$-vertices using Eq. (\ref{Eq:cmt-stationary-1}) where the components $s_l^+$ of the vector $\left|s^+\rangle\right.$ are substituted from Eq. (\ref{Eq:phi-s2}). We have
\begin{align}
&\left[\left(\Omega-i\frac{D_R^TD_R}{2}-\omega\right)+
\frac{1}{2}\sum_{l_n}\gamma_{l_n}\csc\left(kL_{N,l_n}\right) e^{-ikL_{Nl_n}}
 \right]
\phi_N \nonumber\\
&= \frac{1}{2}\sum_{l_n}\gamma_{l_n}\csc\left(kL_{N,l_n}\right)\phi_{l_n}=\frac{1}{2}Y^T\cdot F \cdot \Phi
\label{nlg}
\end{align}

In fact, Eqs. (\ref{hybrid},\ref{nlg}) can be combined and written in a compact form that follows the formulation of the main text (see Eq. (2) in the main text). In our case the matrix ${\tilde M}$ takes the form
\begin{equation}
{\tilde M}=\left(\left[\begin{matrix}
H^{(N-1)}&Y \\
Y^T&H_{NN}
\end{matrix}\right]-iDD^T\right)+\nu(|\phi_N|^2)\left|N\rangle
\langle N\right|
\end{equation}
where $(\left|N\rangle\langle N\right|)_{ij}=\delta_{N,j}$,  and $H_{NN}=2(\omega-\omega_0) +iD_RD_R^T-\sum_{l_n}\left[\gamma_{l_n}\csc(kL_{Nl_{n}})e^{-ikL_{Nl_n}}\right]$.

From Eq. (\ref{hybrid}) we can evaluate the field amplitudes at the vertices of the linear subnetwork. We have
\begin{equation}
\label{FP1}
F\Phi = -\left(H^{(N-1)}-iFDD^TF^T\right)^{-1}\cdot Y \cdot \phi_N\,
\end{equation}
in analogy with Eq. (4) of the main text.

By substituting the expression $F\Phi$ from Eq. (\ref{FP1}) to the right side of Eq. (\ref{nlg}) we come up with the equivalent expression of Eq. (5). Specifically, we have
\begin{widetext}
\begin{equation}
\nu(|\phi_N|^2)=2\left[\left(\omega_0-\omega
-i\frac{D_R^TD_R}{2}\right)+
\frac{1}{2}\sum_{l_n}\gamma_{l_n}\csc\left(kL_{N,l_n}\right) e^{-ikL_{Nl_n}}
 \right]+Y^T\left(H^{(N-1)}-iFDD^TF^T\right)^{-1}\cdot Y
 \label{nlwave}
 \end{equation}
\end{widetext}
which allows us to evaluate the field intensity $|\phi_N|^2$ at the nonlinear resonator (as a function of the incident
wave angular frequency $\omega=ck$). The accepted solutions have to satisfy the following physical conditions
\begin{equation}
0 < |\phi_N|^2 \in \mathbb{R}, \quad k \in \mathbb{R}.
\end{equation}

Equations (\ref{FP1},\ref{nlwave}) are crucial for evaluating the CPA wavefront. Specifically, Eq. (\ref{FP1}) allows us
to evaluate the relative phase and amplitude of the incident wavefront that results in a CPA. For example, if the leads
are attached at a set of vertices $\{\mu\}$, then from continuity at these vertices we get that $I_\mu=\phi_\mu=\langle \mu|F\Phi\rangle\propto \phi_N$ (we always assume that the leads are attached to linear vertices). Therefore, the relative
amplitude and phase of the injected wavefront are independent of the field $\phi_N$ at the nonlinear resonator. Finally,
the amplitude of these injected wavefronts is also evaluated by Eq. (\ref{FP1}) by substituting the magnitude of $\phi_N$
(associated with a CPA frequency) from the solution of Eq. (\ref{nlwave}). The global (undetermined phase) is irrelevant
for the CPA solution.

Up to this point, we did not specify the form $\nu(\cdot)$ of the nonlinear interaction.
There are various options that we can adopt. For example, a Kerr nonlinearity assumes that $\nu(|\phi_N|^2) = \chi_k|\phi_N|^2$, while a saturable nonlinearity assumes that $\nu(|\phi_N|^2) = z_1/[1+\alpha |\phi_N|^2]$ with $\chi_k$, $\alpha$, and $z_1$ being parameters that characterize the specific nonlinear mechanism.
We point out that the above form of the saturable nonlinearity is typical for a variety of wave systems ranging from microwaves\cite{jeon2020non}, optics\cite{Christodoulides97} to acoustics\cite{bourquard_faure21}.
It turns out that the nonlinear diode that we have used is best described by the saturable nonlinearity \cite{jeon2020non} (see last section).

\section{NL-CPA Wavefront Shaping Protocol in Coupled Mode Theory}
In the main text of the paper we have experimentally established an efficient wavefront shaping protocol applicable to achieve CPA using as a testbed, networks of coupled coaxial cables consisting of a single nonlinear vertex. In the previous section of the Supplementary Materials, we have detailed the implemented protocol for the special case of networks.
The developed algorithmic protocol, however, is generic and independent of the specific system framework.
To demonstrate this generality, we turn to Coupled Mode Theory (CMT) to describe a cavity consisting of weakly coupled resonances, of which one, \( n_0 \), is endowed with a known nonlinear mechanism \( \nu\left(|\psi_{n_0}|^2\right) \), where we have used the notation $\psi_{n_0} \equiv \left\langle n_0 \middle| \psi \right\rangle$. The CMT equations read
\begin{subequations}
\begin{equation}
i\frac{d}{dt}|\Psi\rangle = H_{\text{eff}}|\Psi\rangle + \nu|n_0\rangle\langle n_0|\Psi\rangle + iD|S^+\rangle
\end{equation}
\begin{equation}
\left|S^-\right\rangle = D^T\left|\Psi\right\rangle - \left|S^+\right\rangle
\end{equation}
\end{subequations}
The first equation describes internal dynamics of the time-dependent scattering field \( |\Psi\rangle \) in the cavity with respect to a time-dependent incident waveform \( |S^+\rangle \) injected at \( M \) leads whose coupling to the cavity is encoded in the purely real \( N \times M \) coupling matrix \( D \).
The \( N \times N \) effective Hamiltonian \(H_{\text{eff}} = H_0 + \Sigma \) describes dynamics of the linear part of the isolated system together with the self-energy correction due to its coupling with the environment, \(\Sigma = \Delta - i\Gamma \), where the real part \(\Delta \) is associated with a resonant shift and the imaginary part \(\Gamma = \frac{1}{2}DD^T \) describes losses from energy leaking out to the leads [1].
The specific form of \(\Sigma \) depends on the nature of the leads.
Meanwhile, the basis modes of the isolated resonances are denoted by \( |n\rangle \), therefore, the term \( \nu|n_0\rangle\langle n_0| \) indicates the presence of a local nonlinear element associated with the amplitude of the \( n_0^{\text{th}} \) isolated resonance.
The second equation describes continuity of the field at each of the lead ports with respect to the time-dependent outgoing waveform \( |S^-\rangle \).
To develop the NL-CPA wavefront shaping protocol in CMT, we perform a separation of variables on the previous time-domain equations of motion, neglecting higher-order harmonics, \( |\Psi\rangle = e^{-i\omega t}|\psi\rangle \) and \( |S^\pm\rangle = e^{-i\omega t}|s^\pm\rangle \), where \( \omega \) is the driving frequency of the incident wave.
In terms of the time-independent scattering field and wavefronts, we have the nonlinear CMT equations of motion in the frequency domain.
\begin{subequations}
\begin{equation} \label{eq:S2a}
\left(H - \nu|n_0\rangle\langle n_0|\right)
|\psi\rangle= iD|s^+\rangle;\quad H=\omega \mathbb{I}_N
-H_{\text eff}
\end{equation}
\begin{equation} \label{eq:S2b}
|s^-\rangle = D^T|\psi\rangle - |s^+\rangle
\end{equation}
\end{subequations}
where \( \mathbb{I}_N \) is the \( N \times N \) identity matrix.

Similar to the approach taken in the main text, we introduce the \( (N-1) \times N \) filtering matrix \(F \), which serves to extract the linear components of \( |\psi\rangle \), with the following defining properties,
\begin{subequations}
\begin{flalign}
    && F^TF + |n_0\rangle\langle n_0| &\equiv \mathbb{I}_N, \label{eq:S3a} && \\
    && FF^T &\equiv \mathbb{I}_{N-1}, && \\
    && F|n_0\rangle &\equiv 0 \label{eq:S3c} &&
\end{flalign}
\end{subequations}

Explicitly, the elements of the filtering matrix may
be represented, \(F_{ij} = \left\{ \begin{matrix}
\delta_{i,j},\ \ i < n_{0} \\
\delta_{i,j + 1}\ \ i \geq n_{0} \\
\end{matrix} \right.\ \).

We now proceed with formulating the nonlinear wavefront shaping protocol by imposing CPA scattering conditions corresponding to zero outgoing wavefront, \( |s^- \rangle = 0 \) in eq.~\eqref{eq:S2b}.
Inserting the simplified continuity equation into eq.~\eqref{eq:S2a} and re-arranging, we have,
\begin{equation} \label{eq:S4}
{\widetilde M}|\psi\rangle=0;\quad \widetilde{M}\equiv \left( M - \nu|n_0\rangle\langle n_0| \right),
\end{equation}
where \(M=\omega \mathbb{I}_N - \widetilde{H}_{\text{eff}}=H-iDD^T\) and
\( \widetilde{H}_{\text{eff}} \equiv H_{\text{eff}} + iDD^T \) is the effective Hamiltonian with a time-reversed self-energy correction.
Notably, Eq.~\eqref{eq:S4} has an identical form to Eq. (2) in the main text with the only nonlinear component of \(\widetilde{M} \) being its \( n_0-{\text{th}} \) diagonal element, \( \langle n_0 |\widetilde{M} | n_0 \rangle = \omega - \widetilde{\varepsilon}_{n_0} - \nu\left(|\psi_{n_0}|^2\right) \), where \( \widetilde{\varepsilon}_{n_0} \equiv \langle n_0 | \widetilde{H}_{\text{eff}} | n_0 \rangle \), therefore the sub-matrix \(F\widetilde{M}F^T \) contains purely linear elements.
Employing property~\eqref{eq:S3a} of the filtering matrix, we separate the linear and nonlinear parts of Eq.~\eqref{eq:S4},
\begin{equation} \label{eq:S5a}
MF^TF|\psi\rangle =- \psi_{n_0}\widetilde{M}|n_0\rangle
\end{equation}
which is Eq. (3) of the main text.
By multiplying Eq.~\eqref{eq:S5a} by \( \left(F
MF^T \right)^{-1}F \) and re-arranging, we can solve for the linear part of the field in terms of the nonlinear part of the scattering field (see Eq. (4) of the main text),
\begin{equation} \label{eq:S5b}
F|\psi\rangle = -\psi_{n_0} \left(FMF^T \right)^{-1}FM|n_0\rangle
\end{equation}
where on the right-hand side we have again involved property Eq. (\ref{eq:S3c}).
Then, by multiplying Eq.~\eqref{eq:S5a} with the inner product of \( \langle n_0| \) and using property~\eqref{eq:S3c}, we are able to isolate the nonlinear part of the scattering field,
\begin{equation} \label{eq:S5c}
\langle n_0 | M F^T F |\psi\rangle + \psi_{n_0} (\omega - \widetilde{\varepsilon}_{n_0} - \nu) = 0\,.
\end{equation}
Inserting the result for the linear part of the field from eq.~\eqref{eq:S5b} into this expression, and assuming \( \psi_{n_0} \neq 0 \) which cancels, we have the outcome,
\begin{equation}
\langle n_0|MF^T\left( FMF^T \right)^{-1}FM|n_0\rangle + (\omega - \widetilde{\varepsilon}_{n_0} - \nu) = 0\,
\end{equation}
which reproduces Eq. (5) of the main text and can be used to solve for the field intensity corresponding to the nonlinear resonance and, therefore, the field amplitude up to an arbitrary global phase,
\begin{equation} \label{eq:S6}
\nu(|\psi_{n_0}|^2) = \langle n_0 | M F^T \left( F M F^T \right)^{-1} F M | n_0 \rangle + \left( \omega - \widetilde{\varepsilon}_{n_0} \right)\,.
\end{equation}
Therefore, NL-CPA scattering conditions can be identified from a single-parameter search in \( \omega \), whenever the solution for $|\psi_{n_0}|^2$ from Eq.~\eqref{eq:S6} is real and positive.
The result for the nonlinear part of the field can then be re-inserted into Eq.~\eqref{eq:S5b} to obtain the rest of the NL-CPA field.
Lastly, owing to the simplicity of the CPA scattering conditions, the incident NL-CPA wavefront can easily be acquired, \( |s^+ \rangle = D^T |\psi \rangle = D^T F^T F |\psi \rangle + \psi_{n_0} D^T |n_0 \rangle \).

\section{General Solution and Evaluation of Absorbance}
In the previous section, the generic merit of our NL-CPA protocol was demonstrated by justifying its applicability in the abstract framework of CMT.
As it turns out, the notation introduced is also useful to establish an analytic solution to the general scattering problem, irrespective of the scattering conditions (e.g., CPA), for systems consisting of a single local nonlinear element.
To this end, we again proceed by employing properties of the filtering matrix to isolate the nonlinear part of the field.
Inserting Eq.~\eqref{eq:S3a} into Eq.~\eqref{eq:S2a} and re-arranging the terms, we have,
\begin{equation} \label{eq:S7}
\begin{split}
\left( \omega \mathbb{I}_N - H_{\text{eff}} \right) F^T F |\psi\rangle =\\
\quad \psi_{n_0} H_{\text{eff}} |n_0\rangle + \psi_{n_0} (\nu - \omega) |n_0\rangle + i D |s^+ \rangle\,.
\end{split}
\end{equation}
Defining the filtered Green function as \(\widetilde{G} \equiv \left[ F \left( \omega \mathbb{I}_N - H_{\text{eff}} \right) F^T \right]^{-1} \), we multiply Eq.~\eqref{eq:S7} by \( \widetilde{G} F \) and, utilizing property~\eqref{eq:S3c}, we obtain the linear part of the scattering field,
\begin{equation} \label{eq:S8}
F|\psi\rangle = \psi_{n_0} \widetilde{G} F H_{\text{eff}} |n_0\rangle + i \widetilde{G} F D |s^+ \rangle\,.
\end{equation}
Then, by multiplying Eq.~\eqref{eq:S7} by \( \langle n_0| \), we can isolate the nonlinear part of the field,
\begin{equation} \label{eq:S9}
- \langle n_0 | H_{\text{eff}} F^T F | \psi \rangle = \psi_{n_0} (\varepsilon_{n_0} + \nu - \omega) + i \langle n_0 | D | s^+ \rangle\,
\end{equation}
where $\varepsilon_{n_0} \equiv \left\langle n_0 \middle| H_{\text{eff}} \middle| n_0 \right\rangle$.
Inserting the linear part ~\eqref{eq:S8} to this result, we have the succinct relation,
\begin{subequations}
\begin{equation} \label{eq:S10a}
\beta_0 = \psi_{n_0} (\nu + \beta_1)\,,
\end{equation}
where the coefficients are defined,
\begin{equation} \label{eq:S10b}
\beta_0 \equiv -i \langle n_0 | \left( H_{\text{eff}} F^T \widetilde{G} F + \mathbb{I}_N \right) D | s^+ \rangle
\end{equation}
\begin{equation} \label{eq:S10c}
\beta_1 \equiv \langle n_0 | H_{\text{eff}} F^T \widetilde{G} F H_{\text{eff}} | n_0 \rangle - \left( \omega - \varepsilon_{n_0} \right)\,.
\end{equation}
\end{subequations}
Computing the norm-squared of Eq.~\eqref{eq:S10a} and solving for zero, denoting \( I \equiv |\psi_{n_0}|^2 \),
\begin{equation} \label{eq:S11}
I |\nu(I)|^2 + 2I \text{Re} \left[ \beta_1^\ast \nu(I) \right] + I |\beta_1|^2 - |\beta_0|^2 = 0\,.
\end{equation}
Which may be solved using Cardano's formula once an explicit form of the nonlinearity is specified and Eq.~\eqref{eq:S11} is expressed as a cubic polynomial in \( I \),
\begin{equation} \label{eq:S12}
aI^3 + bI^2 + cI + d = 0\,.
\end{equation}
In case of a Kerr-type nonlinearity of the form \( \nu(I) = \chi I \), the coefficients of Eq.~\eqref{eq:S12} are given by,
\begin{subequations}
\begin{equation} \label{eq:S13a}
a = |\chi|^2; \quad b = 2\text{Re}\left[\beta_1^\ast \chi\right]; \quad c = |\beta_1|^2; \quad d = -|\beta_0|^2\,.
\end{equation}
Meanwhile, in case of a saturable-type nonlinearity of the form \( \nu(I) = z_1 - \frac{z_0}{1 + \alpha I} = \frac{\delta + \chi I}{1 + \alpha I} \) with \( \delta \equiv z_1 - z_0 \), \( \chi \equiv z_1 \alpha \) and \( \text{Re}[\alpha] > 0 \), the coefficients of Eq.~\eqref{eq:S12} take the more complicated form,
\begin{equation} \label{eq:S13b}
\begin{gathered}
a = |\chi|^2 + 2\text{Re}[\beta_1^\ast \chi \alpha^\ast] + |\alpha \beta_1|^2, \\
b = 2\text{Re}[\delta^\ast \chi + \beta_1^\ast (\chi + \delta \alpha^\ast)] + 2\text{Re}[\alpha] |\beta_1|^2 - |\alpha \beta_0|^2, \\
c = |\delta|^2 + 2\text{Re}[\beta_1^\ast \delta] + |\beta_1|^2 - 2\text{Re}[\alpha] |\beta_0|^2, \\
d = -|\beta_0|^2\,.
\end{gathered}
\end{equation}
\end{subequations}
Once the coefficients of the cubic polynomial are obtained, Cardano's formula provides the roots \( k \in \{0, 1, 2\} \), via
\begin{subequations}
\begin{equation} \label{eq:S14a}
I_k = -\frac{1}{3a} \left( b + \zeta^k C + \frac{\Delta_0}{\zeta^k C} \right)\,.
\end{equation}
Where \( \zeta \equiv \frac{-1 + i\sqrt{3}}{2} = e^{i\frac{2\pi}{3}} \) is the primitive 3rd root of unity and the constants \( \Delta_0, \Delta_1 \) and \( C \) are provided by,
\begin{equation} \label{S14b}
\begin{gathered}
\Delta_0 \equiv b^2 - 3ac, \\
\Delta_1 \equiv 2b^3 - 9abc + 27a^2d, \\
C \equiv \left( \frac{\Delta_1 \pm \sqrt{\Delta_1^2 - 4\Delta_0^3}}{2} \right)\,.
\end{gathered}
\end{equation}
\end{subequations}
Once physical solutions are recovered, corresponding to real and positive roots \( I_k \), \( \nu(I) \) can be evaluated for the respective incident frequency \( \omega \) and wavefront \( |s^+\rangle \).
With \( \nu \) now known, a conventional linear approach becomes applicable, specifically ~\eqref{eq:S2a} provides the scattering field,
\begin{equation} \label{eq:S15a}
\left| \psi \right\rangle = i G_{\{\omega, |s^+ \rangle\}} D \left| s^+ \right\rangle
\end{equation}
\begin{equation} \label{eq:S15b}
G_{\{\omega, |s^+ \rangle\}} \equiv \left[ \omega \mathbb{I}_N - H_{\text{eff}} - \nu |n_0 \rangle \langle n_0| \right]^{-1}
\end{equation}
Then from Eq.~\ref{eq:S15a}, Eq.~\ref{eq:S2b} provides the corresponding output wavefront,
\begin{equation} \label{eq:S16a}
\left| s^- \right\rangle = S_{\{\omega, |s^+ \rangle\}} \left| s^+ \right\rangle\,,
\end{equation}
where the nonlinear scattering function,
\begin{equation}
S_{\{\omega, |s^+ \rangle\}} \equiv -\mathbb{I}_N + i D^T G_{\{\omega, |s^+ \rangle\}} D
\end{equation}
is dependent on the driving frequency of the incident wave as well as its amplitude and relative phase at each of the ports.
Eventually, the absorbance can be evaluated as:
\begin{equation}
    A=1-\frac{\langle s^-|s^-\rangle}{\langle s^+|s^+\rangle}
\end{equation}
\section{Nonlinear CPA and EP CPA in Ring Graph}
\begin{figure*} [hbt!]
\centering
\includegraphics[width=\linewidth]{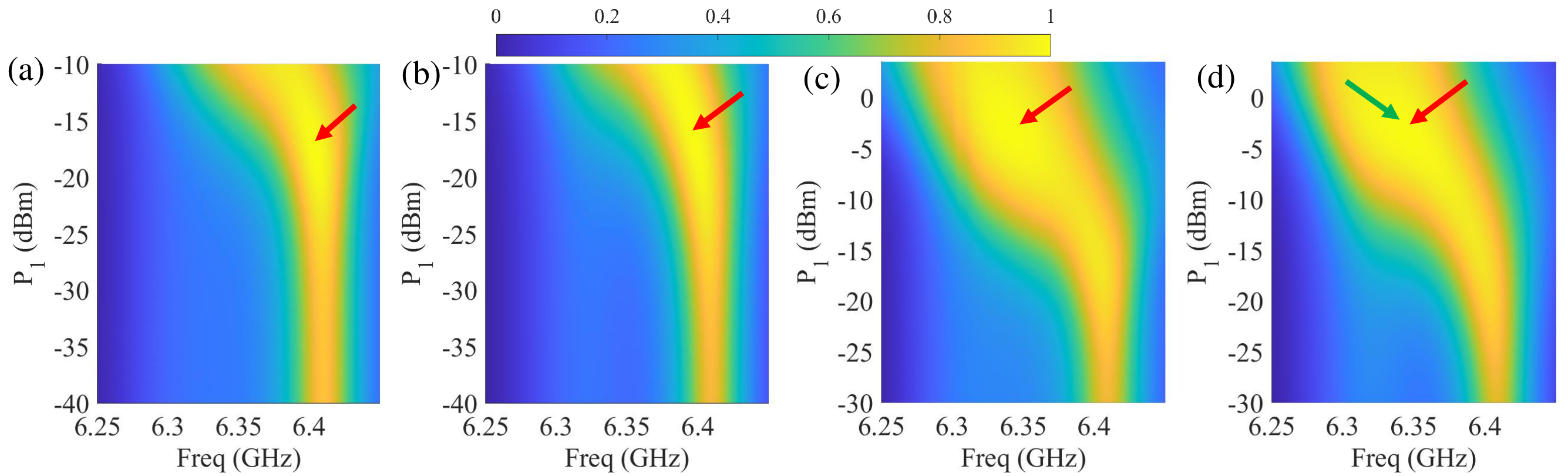}
\caption{ \label{Ring_NL_Compare}
(a) Experimentally measured absorbance as a function of input frequency and power $P_1$ controlled by the IQ modulator 1. The peak absorption is $A^*=99.998$\% occurring at $P_1^*\approx -17.67$\,dBm, $P_2^*\approx -21.09$\,dBm, a relative phase of $\phi^*\approx 97^{\circ}$, and a frequency of $f^*\approx6.391$\,GHz. The experimental wavefront maintained the optimal values of the power ratio $R^*\approx 2.2$ and relative phase $\phi^*\approx 97^{\circ}$.
(b) Theoretically calculated absorbance versus frequency $f$ and input power $P_1$ using a fixed power ratio $R^{cpa}\approx 2.9$ and relative phase $\phi^{cpa}\approx 97^{\circ}$ that corresponds to the optimal wavefront profile evaluated from the NL-CPA protocol. The other theoretical NL-CPA conditions are $P_1^{\text{cpa}}\approx -16.1$\,dBm, $P_2^{\text{cpa}}\approx -20.67$\,dBm, and frequency of $f^{\text{cpa}}\approx 6.397$\,GHz.
(c) Experimentally measured absorbance of a broadband NL-CPA state, versus frequency and input power $P_1$. The peak absorption is $A^*=99.55$\% at frequency $f^*\approx 6.34$\, GHz, with $P_1^*\approx -3.5$\,dBm, $P_2^*\approx -10.90$\,dBm,
and relative phase $\phi_1^*=66^{\circ}$. The associated wavefront used in these measurements is characterized by a fixed power ratio $R^*\approx 5.5$ and relative phase $\phi_1^*\approx 66^{\circ}$.
(d) The same as (b) for a broadband NL-CPA scenario using a wavefront with fixed power ratio $R^{cpa}\approx 6.3$ and relative phase $\phi^{cpa}\approx 120^{\circ}$. The NL-CPA state corresponds to input powers $P_1^{\text{cpa}}\approx -3.6$\,dBm and $P_2^{\text{cpa}}\approx -11.6$\,dBm, and a relative phase of $\phi^{\text{cpa}}\approx 120^{\circ}$ and occur at frequency $f^{\text{cpa}}\approx 6.34$\,GHz.
The CPA point in all subfigures is marked with a red arrow. In subfigure (d) we indicate with an additional green arrow the position of the second (case (b) of main text) NL-CPA.
}
\end{figure*}
In Fig.~\ref{Ring_NL_Compare}, we present experimental and theoretical plots of absorption as a function of frequency
and input power. Specifically, Fig.~\ref{Ring_NL_Compare}(a) shows the experimental results for narrow-band CPA, for
fixed injected power ratio $R^*\approx 2.2$ and relative phase $\phi^*\approx 97^{\circ}$-- see also
Fig.~2(a) of the main text. The associated numerical calculation of absorbance (see Fig.~
\ref{Ring_NL_Compare}(b)) uses a wave with fixed injected power ratio $R^{cpa}\approx 2.9$ and relative phase $\phi^{cpa}
\approx 97^{\circ}$-- see also main text -- associated with the NL-CPA wavefront that was evaluated using our protocol.

Similarly, Fig.~\ref{Ring_NL_Compare}(c) is the experimental results for a wide-band CPA, where the injected wavefront
has fixed injected power ratio $R^*\approx 5.5$ and relative phase $\phi^*\approx 66^{\circ}$-- see also Fig.~2(b) of the main text. The
analogous theoretical analysis is shown in Fig.~\ref{Ring_NL_Compare}(d), demonstrating a similar behavior as our experimental
data. The calculation has been performed using a fixed injected relative power $R^{cpa}\approx 6.3$, and relative phase $\phi^{cpa}\approx 120^{\circ}$. These values correspond to the optimal wavefront -- see also main text -- and have been
evaluated using our NL-CPA protocol.

\begin{figure*} [hbt!]
\centering
\includegraphics[width=\linewidth]{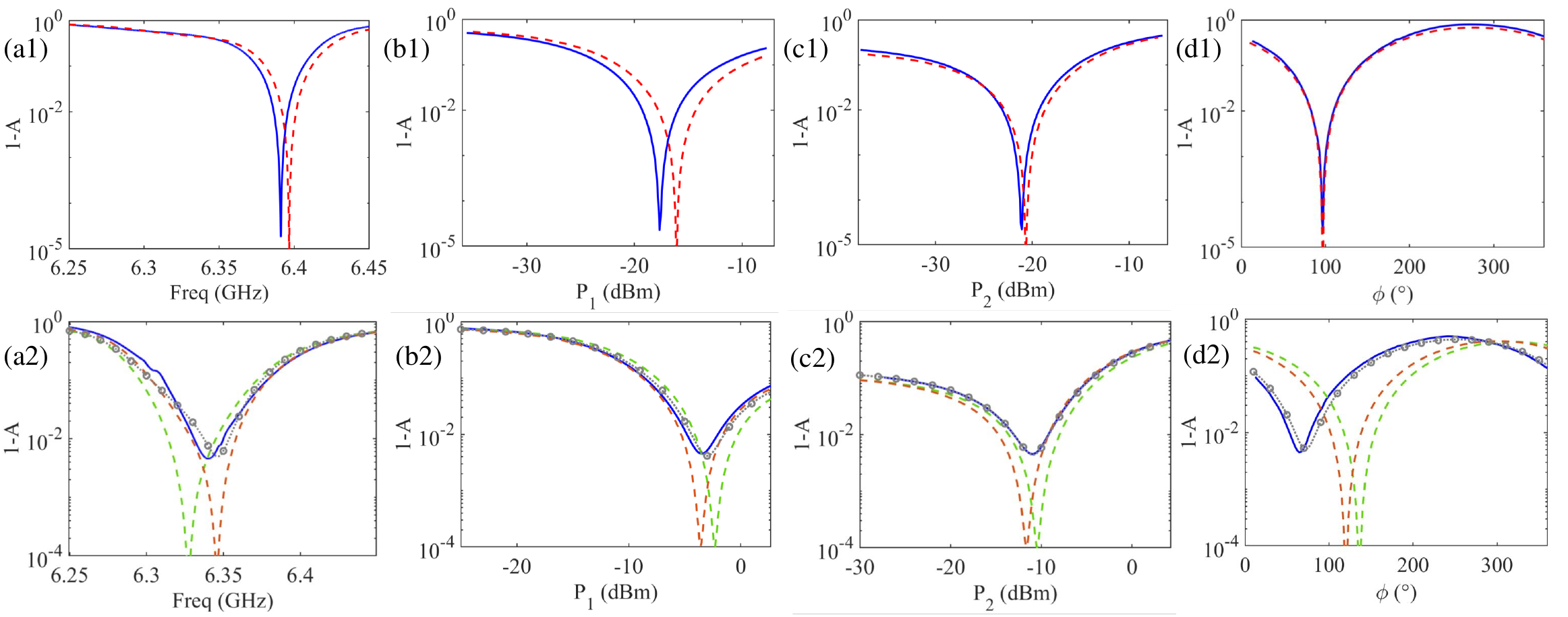}
\caption{ \label{Ring_Para_Scan}
(a1)-(d1) Outgoing signal versus frequency (a1), input power $P_1$ (b1), input power $P_2$ (c1), and relative phase $\phi_1$ (d1) for the narrow-band CPA. The solid blue lines are the experimental measurements, while the red dashed lines are the theoretical results.
For each plot, we fix the three remaining wavefront parameters at the NL-CPA conditions (see figure caption of Fig.~\ref{Ring_NL_Compare}a-b).
(a2)-(d2) The same analysis as in the upper row, but now for the broadband NL-CPA scenario. The red and green dashed-lines correspond to different wavefront (fixed) parameters associated with each (quasi-degenerate) NL-CPAs (see green and red arrows in Fig.~\ref{Ring_NL_Compare}d) that have been found using our NL-CPA protocol (see text). The grey-dashed line
with circle markers corresponds to the wavefront of the case (a) of the main text in case the lengths of
the ring graph has varies by less than $\pm 2$mm.
}
\end{figure*}
\begin{figure*}
    \centering
    \includegraphics[width=\linewidth]{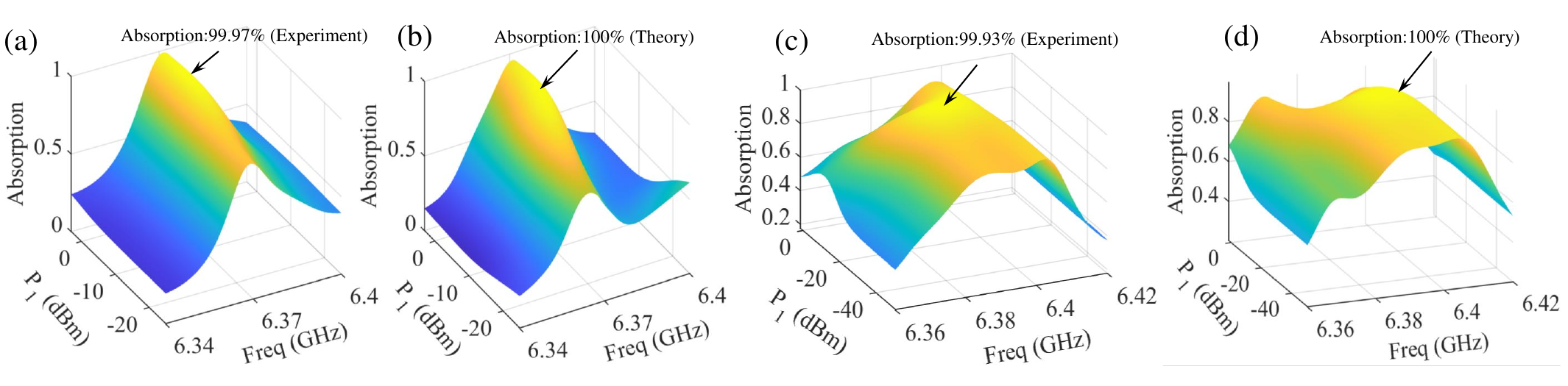}
    \caption{ \label{Absorption_Power_Tetra}
        (a) Experimentally measured absorption profile for narrow-band NL-CPA as a function of input frequency and power $P_1$ from IQ modulator 1, achieving a peak absorption of 99.97\%.
        The corresponding parameters include $P_1^{\text{cpa}}=-5.09$\,dBm, $P_2^{\text{cpa}}=-12.90$\,dBm, a relative phase of $\phi_1^{\text{cpa}}=207^{\circ}$, and a frequency of $f^{\text{cpa}}=6.371$\,GHz.
        It should be noted that we maintain a constant ratio between $P_1$ and $P_2$ as the power is varied.
        (The CPA point in this subfigure, as well as in subsequent subfigures are marked with black arrows.)
        (b) Theoretically calculated absorption profile for narrowband NL-CPA using the hybrid model, achieving a peak absorption of 100\%.
        The theoretical NL-CPA conditions are characterized by $P_1^{\text{cpa}}=-6.6$\,dBm, $P_2^{\text{cpa}}=-14.35$\,dBm, a relative phase of $\phi_1^{\text{cpa}}=191^{\circ}$, and a frequency of $f^{\text{cpa}}=6.369$\,GHz.
        (c)  Experimentally measured absorption profile of a broadband NL-CPA state, where the peak absorption recorded is 99.93\%.
        The associated wavefronts have $P_1^*\approx -19.05$\,dBm, $P_2^*\approx -15.6$\,dBm, a relative phase of $\phi_1^*\approx301^{\circ}$, and a frequency $f^*\approx 6.39$\,GHz.
        (d) Theoretical analysis of the broadband NL-CPA state, where the predicted peak absorption achieves 100\%, with the corresponding input powers as $P_1^{\text{cpa}}=-19.24$\,dBm and $P_2^{\text{cpa}}=-17.79$\,dBm, a relative phase of $\phi_1^{\text{cpa}}=377.8^{\circ}$, and a frequency of $f^{\text{cpa}}=6.397$\,GHz.
        }
\end{figure*}

To confirm the existence of NL-CPA and its corresponding wavefront, we analyzed the outgoing relative signals
${\cal O}=1-A$ against each of the remaining parameters, fixing the rest of them at their NL-CPA values.
Figures~\ref{Ring_Para_Scan}(a1)-(d1) show the experimental measurements (blue solid line) of ${\cal O}$ versus
frequency $f$, input powers $P_1$ and $P_2$ from channels 1 and 2 (IQ-modulator 1 and 2, respectively), and the
relative phase $\phi$ between channel 1 and channel 2, respectively. At the same figures, we also report the
theoretical results (red dashed lines) derived using the two NL-CPA wavefronts predicted by our protocol. The
analogous parametric analysis for the broadband NL-CPA scenario is illustrated in Figs.~\ref{Ring_Para_Scan} (a2)-(d2).
The experimental deeps (demonstrating optimal CPA conditions) typically fall within the parameter domain where
the two NL-CPAs are predicted (red and green dashed lines corresponding to cases (a) and (b), respectively, in
the main text). The relative phase discrepancy between the experimental
and the theoretical results, where the maximum absorption occurs, is associated with the measurement precision
of the measured lengths. For example, a small variation of the lengths of the network by less than $\pm 2$mm
can shift the emitted signal dip close to the experimental findings without affecting the nice agreement of
the other optimal wavefront parameters (see grey-dotted lines with circle markers in Fig.~\ref{Ring_Para_Scan}
(a2)-(d2) where we have used the CPA wavefront associated with the case (a) of the main text ).

\section{NL-CPA and zero-EPDs in Tetrahedron Graph}
We have repeated the same analysis as above for the case of a tetrahedron graph. Two graph configurations \cite{SM}
have been used such that a narrow-band and a broad-band NL-CPA can be examined in detail within
the range of frequency and power variation that is available by our VNA.

The first configuration depicts a narrow-band NL-CPA scenario and it is shown in Fig. \ref{Absorption_Power_Tetra}(a) (experiment) and Fig. \ref{Absorption_Power_Tetra}(b) (theory). In each of these two figures, we kept the power ratio $R$ and relative phase $\phi$ of the injected wavefront fixed and varied the $P_1$ and the injected frequency $f$. The fixed parameters $R,\phi$ have been chosen to match the ones corresponding to the maximum absorption scenario ($R^*\approx 6.03$ and $\phi^*\approx 207^{\circ}$ with maximum absorption $A^*=99.97\%$ for the experiment and $R^{\text cpa}\approx 5.96$ and $\phi^{\text cpa}\approx 191^{\circ}$ for the theory). The narrow-band NL-CPA occurs at $P_1^*\approx -5.09$dBm and frequency $f^*\approx 6.371$GHz for the experiment and $P_1^{\text cpa}\approx -6.6$dBm and $f^{\text cpa}\approx 6.369$GHz for the theory. The comparison between the theoretical and the experimental results confirms the efficiency of our NL-CPA protocol and establishes the fact that the injected power can be used as an external tuning knob for achieving CPA. The small differences between theoretical and experimental parameters are associated with the accuracy of the extraction of the experimental parameters (electrical lengths of the graph and the T-junctions, evaluation of the electrical permittivity etc.).

The graph configuration in Fig.~\ref{Absorption_Power_Tetra}(c), depicts a broad-band NL-CPA scenario which again highlights power-dependent absorption, peaking at $A^*\approx 99.93\%$. The corresponding theoretical results are shown in Fig.~\ref{Absorption_Power_Tetra}(d) and agree well with the observed experimental data. We remind that in this case, our protocol has been identified two (quasi-degenerate) NL-CPAs (see Fig. 4(b) and related discussion in the main text). As previously, the wavefront used in these cases maintains a fixed injected power ratio $R$ and relative phase $\phi$ corresponding to the values associated with the optimal configuration: for the experimental analysis we used $R^*\approx 0.45$ and $\phi^*\approx 301^{\circ}$ while for the theoretical calculations, we have used the parameters associated with the NL-CPA of case (a) of the main text which corresponds to $R^{\text cpa}\approx 0.68$ and $\phi^{\text cpa}\approx 377.8^{\circ}$.

\begin{figure*} [hbt!]
\centering
\includegraphics[width=\linewidth]{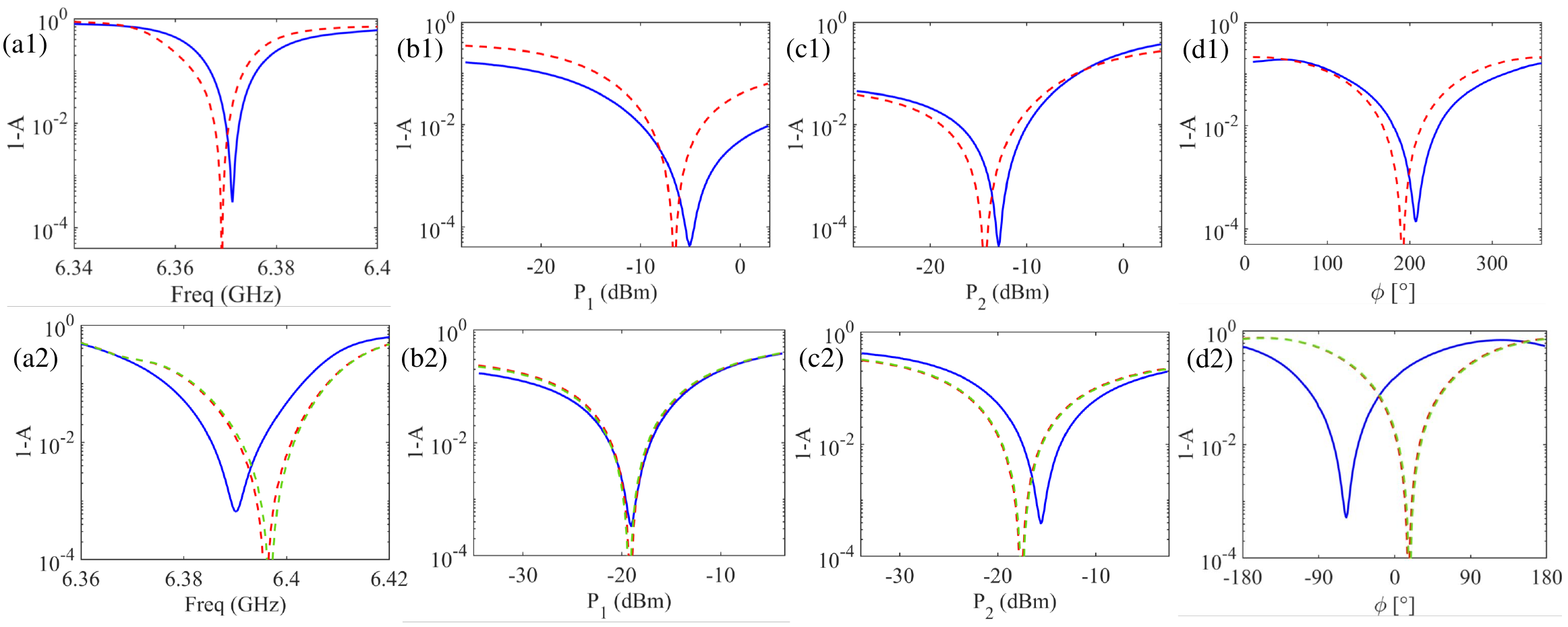}
\caption{ \label{Tetra_Para_Scan}
Relative outgoing signal
(${\cal O}=1-A$) versus frequency (a1,a2), input power $P_1$ (b1,b2), input power $P_2$ (c1,c2), and relative phase
$\phi_1$ (d1,d2) for the narrow-band (upper row) and broadband (lower row) CPA respectively. For each plot, we fix the
three remaining parameters taken to be at the NL-CPA conditions (see text). The solid blue lines are the experimental
results while the dashed lines are the NL-CPA wavefronts predicted by our algorithm. The green and red dashed lines in
the lower row are the theoretical results for the absorbance evaluated using two different wavefronts (case (a) and
case (b) of the main text) associated with each (quasi-degenerate) NL-CPA that have been found using our algorithm. }
\end{figure*}


To confirm the existence of the NL-CPAs and their corresponding wavefront against the predictions of our protocol, we analyzed the relative outgoing signal ${\cal O}=1-A$ against each of the four parameters, maintaining the remaining parameters at their NL-CPA values. Figs.~\ref{Tetra_Para_Scan}(a)-(d) show the experimentally measured ${\cal O}$ (blue solid lines) versus frequency $f$, input powers $P_1$ and $P_2$ and the relative phase $\phi$ of the injected waves, from TLs 1 and 2, respectively. At the same figures, we also report the theoretically evaluated outgoing relative signal ${\cal O}$ using the parameters (see main text) for each of the two NL-CPAs predicted by our algorithm (red and green dashed lines). Due to the proximity of the two NL-CPAs there is hardly visible any difference between them. The excellent agreement between theoretical results and experimental data reconfirms the applicability of our NL-CPA algorithm to predict the appropriate optimal wavefronts that lead to extreme absorption.

\begin{figure}
    \centering
    \includegraphics[width=0.55\linewidth]{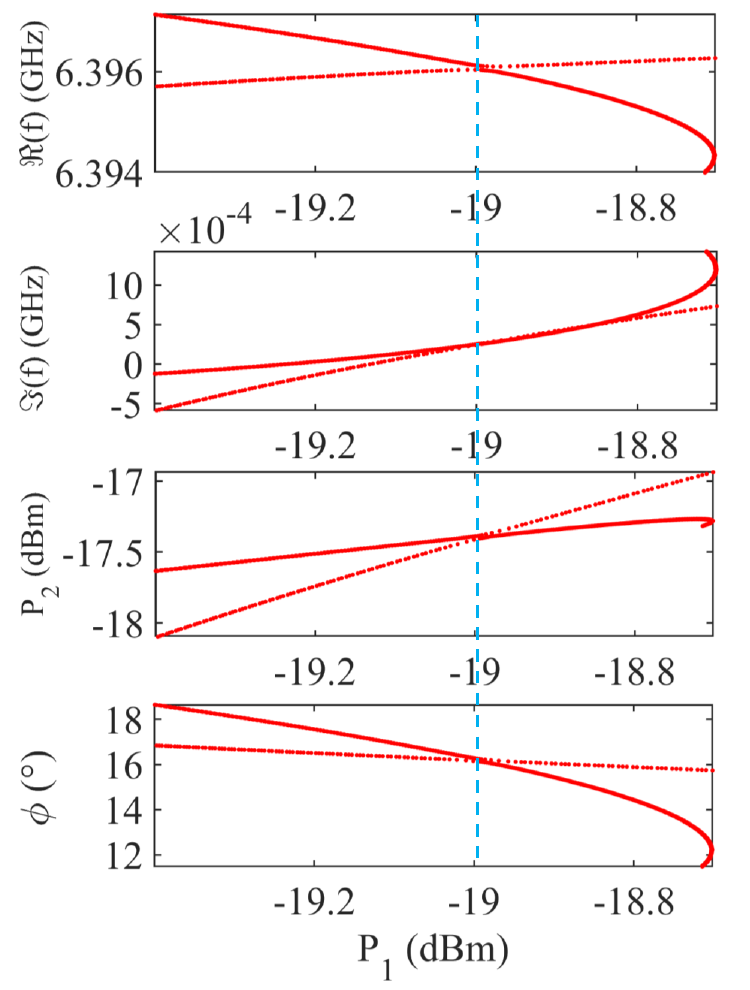}
    \caption{ \label{Degeneracy_Tetra}
        The same plot as in Fig.~2(d) of the main text. The sub-figures demonstrate the dependence of the theoretically calculated wavefront parameters associated with a scattering zero, versus the input intensity from channel 1 ($P_1 = |I_1|^2$): (a) the real part of frequency ($\Re(f)$), (b) imaginary part of frequency ($\Im(f)$), (c) input wave intensity from channel 2 ($P_2 = |I_2|^2$), and (d) relative phase $\phi$ between the injected waves at the two TLs. The dashed (solid) lines indicate the same zero-mode in all subfigures. These modes, coalesce at the same parameter values (light blue dashed line) forming an NL-EPD. Notice that the formed EPD is displaced from the real frequency axis by ${\cal I}(f)\sim 10^{-4}$.
        }
\end{figure}

We have scrutinized further the complex-zero EPD by analyzing its dependence on the input wavefront parameters i.e. their real and imaginary frequency, injected power $P_2$, and relative phase of the incident waves at TLs 1 and 2, versus the incident power $P_1$ from the TL 1. These results are shown in Figs.~\ref{Degeneracy_Tetra}(a-d) and illustrate a degeneracy of the two zero modes at the same $P_1$ value. This EPD is displaced from the real frequency axis by ${\cal I}(f)\sim 10^{-4}$, which for all practical purposes, is insignificant. As a result, we were able to witness a super-quadratic scaling ${\cal O}\sim (\delta f)^{\alpha}$ with a best fitting power $\alpha= 3.5$ which resembles the expected quartic line-shape of EPD-CPA \cite{sweeney2019perfectly} (see Fig. 4(c) of the main text).

\section{Experimental Implementation of the Lossy Non-Linear Vertex}
The nonlinear vertex consists of a cylindrical resonator (ceramics ZrSnTiO with permittivity $\epsilon\approx
36$, height 5\,mm, diameter 8\,mm, resonance frequency around $f_0^{(R)} \approx 6.885$\,GHz and a line width
$\gamma \approx 1.7$\,MHz) which is inductively coupled to a metallic ring (diameter 3\,mm) that is short-
circuited to a diode (detector Schottky diode SMS 7630-079LF, from Skyworks), see inset of Fig.~1(a) of the main text.

The coupling between the ring and the resonator occurs via the resonator's $z$-directional magnetic field
component that is excited by the injected signal. The latter induces a current at the ring and subsequently
a voltage across the diode whose magnitude is controlled by the magnitude of the magnetic field. The magnitude
of this voltage defines the state of the diode: the “on” state is associated with high voltage (high incident
field power) and leads to high nonlinearities; the “off” state is associated with low voltage (low incident
power) and leads to low nonlinearities. The nonlinear resonator is designed to operate at 6.1-6.5\,GHz.

The resonator-ring structure is coupled with the rest of the graph via three ``kink'' antennas [see Fig.~1(b)]. This nonlinear construction has been already used to realize topological limiters
in a coupled resonator waveguide array \cite{jeon2020non}, while more recently it has been used to analyze
asymmetric transport due to nonlinearities \cite{WKKK23}.

\section{Modeling and characterization of the Lossy Non-Linear Vertex}
We have characterized the nonlinear resonator (i.e.\ the form of the saturable nonlinearity $\nu(|\phi_N|^2)$)
and its coupling constants with three kink antennas by comparing the transmission measurements with the
corresponding expressions from coupled-mode theory that describes a three-port scattering
set-up (see inset in Fig.~1a). Below we follow the methodology used in and also Ref. \cite{WKKK23}.

Combining Eqs.~(\ref{sssf}) together, we get the $3\times 3$ scattering parameters
\begin{align} \label{Sres}
S = -1 + iD_R^T\frac{1}{\omega - \Omega+\frac{i}{2}D_RD_R^T}D_R
\end{align}
where $\Omega=2\pi \left(f_0 + z_0 - \frac{z_1}{1+\alpha |\phi_N|^2}\right)$ and $f_0\approx 6.4$GHz is
the linear resonant frequency of the resonator-ring system. The constant $z_0-z_1$ accounts mainly for
losses (radiative and/or Ohmic) due to the presence of the ring which might be different for weak and
strong input powers. From the experimental point of view, two of the TLs have been connected with the VNA
while in each measurement the third port is coupled to a 50\,Ohm terminator.

For weak input powers (e.g. $\sim-25$dBm), $ \alpha |\phi_N|^2 \approx 0$ and the transmission from TL $m$
to $n$ $(m,n=1,2,3)$ takes the form
\begin{align} \label{Eq:fitting-weak-Snm}
  |S_{nm}|^2 = \frac{4\gamma_n^2\gamma_m^2}{[4\pi(f - f_0 - \Re(z))]^2 + [\sum_{l=1}^3\gamma_l^2 - 4\pi\Im(z)]^2}\,,
\end{align}
where $z = z_0 - z_1$. The maximum value of $|S_{nm}|^2$ occurs at $f_{\rm max} = f_0
+ \Re(z)$. Therefore, the measurement of $|S_{nm}(f_{\rm max})|^2$ can be used for extracting $f_0$ (we
consider $\Re(z)= 0$ for simplicity)
together with the $\Im(z) = \frac{\sum_{l=1}^3\gamma_l^2 }{4\pi} \pm \frac{\gamma_n \gamma_m}{2\pi|S_{nm}(f_{\rm max})|}$. Substituting these into Eq.~(\ref{Eq:fitting-weak-Snm}), allows us to express the scattering parameters in terms of $\gamma_l$'s. The latter
are extracted via a direct fitting with the measured $|S_{nm}(f)|^2$ versus $f$. This analysis allows us to evaluate also $\Im(z)$ which satisfy the constraint $\Im(z)\leq \frac{\left(\sum_{l=1}^3\gamma_l^2\right) - \gamma_n^2} {4\pi}$. The latter bound is enforced by the requirement that the reflectance, which
in the weak incident power limit takes the form,
\begin{widetext}
\begin{align} \label{refl}
|S_{nn}|^2 = 1 - \frac{4\gamma_n^2\left[\left(\sum_{l=1}^3\gamma_l^2 \right) - \gamma_n^2 - 4\pi\Im(z)\right]
}{[4\pi(f - f_0 - \Im(z))]^2 + \left[\left(\sum_{l=1}^3\gamma_l^2 \right) - 4\pi\Im(z)\right]^2}
\end{align}
\end{widetext}
must be bounded from above by unity.

The extraction of $z_0$ is achieved by performing a similar analysis of the transmission and reflection spectra for strong input powers. Since now
$\frac{\tilde{z}_1}{1+\chi_s|a|^2} \approx 0$, Eqs.~(\ref{Eq:fitting-weak-Snm},\ref{refl}) will need to be modified by substituting $z\rightarrow z_0$. By
repeating the same procedure as previously, we extract $z_0$ which together with the previous knowledge about $z = z_0 - z_1$ allows us to get also $z_1$. Finally, the nonlinear coefficient $\alpha$ has been estimated by a best-fitting analysis for a set of experimental scattering data that we have collected for intermediate values of incident power.

The above analysis led to the following best fitting values for
the saturable nonlinearity: $z_0 = (-87.5 - 40.8536i)$ MHz, $z_1 = (-87.5 - 32.5189)$ MHz, $\gamma_1^2=89.5$ MHz,$\gamma_2^2=88.5$ MHz,$\gamma_3^2 = 64.7$ MHz,
$\alpha = (3+1.5i)\cdot 10^{9}$(mW$\cdot$s)$^{-1}$. And we should note that, we adjusted these parameters for the ring and tetrahedron graph fittings in order to get relatively better fitting for specific configuration and frequency ranges.
\begin{figure}
    \centering
    \includegraphics[width=0.75\linewidth]{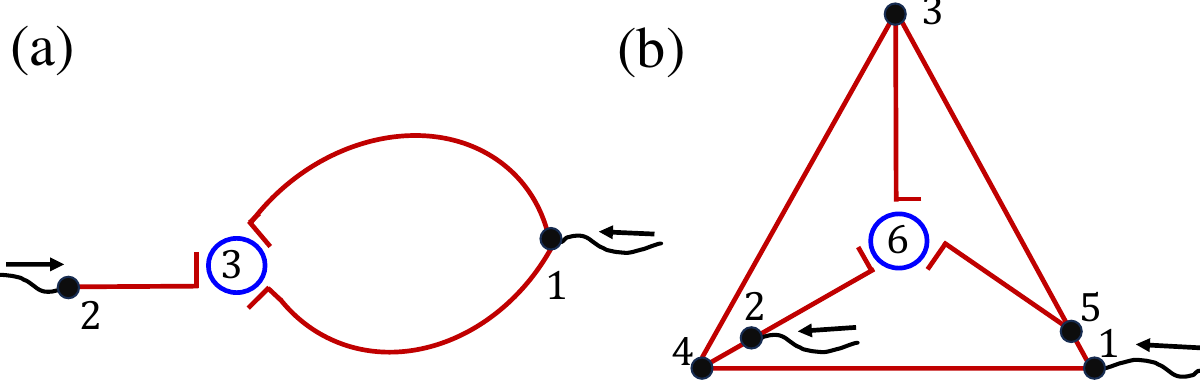}
    \caption{ \label{Fig:schematics}
        Schematics of (a) ring graph and (b) tetrahedron graph.
        }
\end{figure}

\section{Experimental configuration of ring network}
For the ring network sketched in Fig.~\ref{Fig:schematics} (a), the upper arm cable of the ring has a geometric length  $L^{\text{upper}}_{13} = 291$\,mm, while the lower arm cable has length $L^{\text{lower}}_{13} =339$\,mm. The left cable antenna has length $L_{23} = 277$ mm. Note that in each arm length we have included an additional geometric length of $9$mm that estimates the effective length introduced by the T-junction. The measurement precision for the cable lengths is $\pm 2$mm. The extracted cable refractive index is $n=1.212+0.001i$.

\noindent{\bf Theoretical modeling parameters for ring network --}. The best-fitting parameters for the ring network configuration were identified by comparing our experimental scattering parameters with the results from the modeling for the frequency range of interest. We extracted $L^{\text{upper}}_{13} = 286.23$\,mm, $L^{\text{lower}}_{13} =319$\,mm, and $L_{23} = 282.2$ mm, with cable refractive index $n=1.212+0.001i$. We have further refined the nonlinear resonator parameters (see previous section) such that $f_0=6.3933$ GHz, $z_0 = -87.5-41.5943i$ MHz, $z_1 = -87.5 - 33.6641i$ MHz, $\alpha = (2+1.5i)\cdot 10^9$(mW$\cdot s)^{-1}$, $\gamma_1^2 = 78.4$ MHz, $\gamma_2^2 = 89.2$ MHz, $\gamma_3^2 = 82.8$ MHz.

\noindent{\bf Experimental configuration of tetrahedron network --}. The tetrahedron network shown in Fig.~\ref{Fig:schematics}(b) is characterized by the following geometric length configuration: $L_{14}=770$ mm, $L_{15}=18$ mm, $L_{24}=18$ mm, $L_{26}=328$ mm, $L_{34}=289$ mm,  $L_{36}=339$ mm, $L_{56}=291$ mm, $L_{35}=942$ mm (including also an estimation for the length of the phase shifter), with cable refractive index $n=1.212+0.001i$. Note that in each $L_{ij}$ we have included an additional geometric length of $9$mm that estimates the effective length introduced by any T-junction attached to a cable. The measurement precision for the cable lengths is $\pm 2$mm.

\noindent{\bf Theoretical modeling parameters for tetrahedron network -- }. Following the same procedure as for the modeling of the ring network we have identified $L_{14}=760.2$ mm, $L_{15}=20.2$ mm, $L_{24}=20.2$ mm, $L_{26}=341.2$ mm, $L_{34}=309$ mm,  $L_{36}=334.4$ mm, $L_{56}=291.9$ mm, $L_{35}=969.6$ mm, with cable refractive index $n=1.212+0.001i$. We have further refined the nonlinear resonator parameters (see the previous section) such that $f_0=6.3933$ GHz, $z_0 = -70.73-35.047i$ MHz, $z_1 = -70.73 - 27.46i$ MHz, $\alpha = (0.7558+1.411i)\cdot 10^9$(mW$\cdot s)^{-1}$, $\gamma_1^2 = 82.0$ MHz, $\gamma_2^2 = 78.0$ MHz, $\gamma_3^2 = 62.5$ MHz. We have also allowed for small variations between these parameters and the one used for narrow-band CPA modeling.

\end{document}